\documentclass[10pt,journal,compsoc]{IEEEtran}
\usepackage{multirow}
\usepackage{footmisc}
\usepackage{subcaption,siunitx,booktabs}
\usepackage{algorithmic}
\usepackage{caption}
\usepackage{bbm}
\usepackage{todonotes}
\usepackage{amsmath}
\usepackage{amsthm}
\usepackage{textcomp}
\usepackage{xcolor}
\usepackage{amsmath,amssymb,amsfonts}

\usepackage{balance}
\usepackage{enumitem}
\usepackage[ruled]{algorithm2e}
\usepackage{url}
\definecolor{mygray}{gray}{0.9}
\definecolor{comment}{RGB}{0, 102, 0}
\definecolor{darkolivegreen}{rgb}{0.33, 0.42, 0.18}
\definecolor{darkpastelgreen}{rgb}{0.01, 0.75, 0.24}
\definecolor{mediumjunglegreen}{rgb}{0.11, 0.21, 0.18}

\SetKwInput{KwInput}{Input}                
\SetKwInput{KwOutput}{Output}              
\SetKwComment{Comment}{$\triangleright$\ }{}
\SetKwFor{For}{for (}{) $\lbrace$}{$\rbrace$}

\theoremstyle{definition}

\usepackage{etoolbox}

\makeatletter
\patchcmd{\@algocf@start}
  {-1.5em}
  {0pt}
  {}{}
\makeatother

\renewcommand{\todo}[1]{\iffalse #1 \fi{\color{blue} \textbf{[TODO]}}}
\renewcommand{\todo}[1]{\iffalse #1 \fi}
\newcommand{\done}[1]{\iffalse #1 \fi}

\ifCLASSOPTIONcompsoc
  \usepackage[nocompress]{cite}
\else
  \usepackage{cite}
\fi
\ifCLASSINFOpdf
\else
\fi
\begin{document}
%
\title{Causal Distillation for Alleviating Performance Heterogeneity in Recommender Systems}

\definecolor{myblue}{rgb}{0, 0, 0}
\newcommand{\etal}{\textit{et al}.}
\newcommand{\ie}{\textit{i}.\textit{e}.}
\newcommand{\eg}{\textit{e}.\textit{g}.}
\newcommand{\cf}{\textit{cf}. }
\newcommand{\wrt}{\textit{w}.\textit{r}.\textit{t}. }
\newcommand{\vpara}[1]{\vspace{0.05in}\noindent\textbf{#1 }}
\newcommand\norm[1]{\left\lVert#1\right\rVert}
\newcommand{\zsy}[1]{\textcolor{myblue}{#1}}

%
%
%
%

\author{Shengyu Zhang†, Ziqi Jiang†, Jiangchao Yao, Fuli Feng, Kun Kuang*, Zhou Zhao, Shuo Li, \\ Hongxia Yang, Tat-seng Chua, Fei Wu*,~\IEEEmembership{Senior Member,~IEEE}
	
	\IEEEcompsocitemizethanks{\IEEEcompsocthanksitem Shengyu Zhang, Ziqi Jiang, and Zhou Zhao are with Zhejiang University, Hangzhou 310027, China. E-mail: \{sy\_zhang, jiangzq, zhouzhao\}@zju.edu.cn}%
	\IEEEcompsocitemizethanks{\IEEEcompsocthanksitem Kun Kuang is with Zhejiang University, Hangzhou 310027, China, and also with the Key Laboratory for Corneal Diseases Research of Zhejiang Province, Hangzhou 310016, China. E-mail: \{kunkuang\}@zju.edu.cn%
	\IEEEcompsocthanksitem Fei Wu is with the College of Computer Science and Technology, Zhejiang University, Hangzhou 310027, China,  with the Shanghai Institute for Advanced Study, Zhejiang University, Shanghai 201210, China, and also with the Shanghai AI Laboratory, Shanghai 200232, China.\protect\\ E-mail: {wufei@zju.edu.cn}%
	\IEEEcompsocthanksitem Jiangchao Yao is with the Cooperative Medianet Innovation Center, Shanghai Jiao Tong University, Shanghai 200240, China, and with the Shanghai AI Laboratory, Shanghai 200232, China.
	\IEEEcompsocthanksitem Fuli Feng is with the University of Science and Technology of China, Hefei 230026, China. E-mail: 	fulifeng93@gmail.com
	\IEEEcompsocthanksitem Shuo Li and Tat-Seng Chua are with the National University of Singapore, Singapore. E-mail: 	li.shuo980@gmail.com, chuats@comp.nus.edu.sg
	\IEEEcompsocthanksitem Hongxia Yang is with Alibaba Group, Hangzhou 310052, China. \protect\\ E-mail: yang.yhx@alibaba-inc.com}%
	\thanks{† Equal contribution}
	\thanks{* Corresponding authors}
}

\IEEEtitleabstractindextext{%
\begin{abstract}
Recommendation performance usually exhibits a long-tail distribution over users --- a small portion of head users enjoy much more accurate recommendation services than the others.
We reveal two sources of this performance heterogeneity problem: the uneven distribution of historical interactions (a natural source); and the biased training of recommender models (a model source). 
As addressing this problem cannot sacrifice the overall performance, a wise choice is to eliminate the model bias while maintaining the natural heterogeneity. The key to debiased training lies in eliminating the effect of confounders that influence both the user's historical behaviors and the next behavior. The emerging causal recommendation methods achieve this by modeling the causal effect between user behaviors, however
potentially neglect unobserved confounders (\eg, friend suggestions) that 
are hard to measure in practice. 

To address unobserved confounders, we resort to the front-door adjustment (FDA) in causal theory and propose a causal multi-teacher distillation framework (CausalD). FDA requires proper mediators in order to estimate the causal effects of historical behaviors on the next behavior. To achieve this, we equip CausalD with multiple heterogeneous recommendation models to model the mediator distribution.
Then, the causal effect estimated by FDA is the expectation of recommendation prediction over the mediator distribution and the prior distribution of historical behaviors, which is technically achieved by multi-teacher ensemble.
To pursue efficient inference, CausalD further distills multiple teachers into one student model to directly infer the causal effect for making recommendations.
We instantiate CausalD on two representative models, DeepFM and DIN, and conduct extensive experiments on three real-world datasets, which validate the superiority of CausalD over state-of-the-art methods. Through in-depth analysis, we find that CausalD largely improves the performance of tail users, reduces the performance heterogeneity, and enhances the overall performance.
\end{abstract}

\begin{IEEEkeywords}
Recommender System; Performance Heterogeneity; Causal Distillation; Front-door Adjustment
\end{IEEEkeywords}}

\maketitle

\IEEEdisplaynontitleabstractindextext

%
\IEEEpeerreviewmaketitle

\section{Introduction}

\todo{[DONE]++*

1. The motivation of removing unobserved confounders is not clear in the manuscript. The authors argued that the confounders are unobserved but they only introduced popularity as a confounder. However, popularity is observed, and there are previous works to solve the problem of observed popularity biases [1,2]. A reasonable example is necessary to explain why it is necessary to remove effects from unobserved confounders. My suggestion is to introduce some other clear examples to show there are unobserved but important confounders.

[1] Wei T, Feng F, Chen J, et al. Model-agnostic counterfactual reasoning for eliminating popularity bias in recommender system[C]//Proceedings of the 27th ACM SIGKDD Conference on Knowledge Discovery & Data Mining. 2021: 1791-1800.
[2] Zhao Z, Chen J, Zhou S, et al. Popularity bias is not always evil: Disentangling benign and harmful bias for recommendation[J]. arXiv preprint arXiv:2109.07946, 2021.

Thank you for your valuable feedback. We would like to respectfully inform you that we have included an example of "friends' suggestions/recommendations on items" in our introduction. It is worth noting that the reason behind a user's click, whether it is due to their inherent interests or suggestions from their friends and family, remains unobserved for the model. Social relationships are known to have a significant impact on recommender systems while social recommendation is a well-established research field.

Taking your suggestions into account, we have made further clarifications in both the Introduction and Method sections to provide a clearer understanding of our study. Additionally, we have included an additional example to highlight the importance of unobserved confounders in our research. To support our claims, we have also cited the two papers mentioned in the Related Works section.

}

\todo{[DONE]++*

4. I also have concern about why it is necessary to provide similar recommendation quality for active users and inactive users. Active users provide more interaction data. Therefore, it is natural that recommender systems can better mine their potential interest based on more data, leading to better recommendation quality. On the opposite, inactive users provides less data, it seems natural that inactive users could not obtain the same recommendation quality as active users.

We understand your concern on the motivation clarification of this paper. We agree that active users should receive better recommendation quality due to better interest mining over more interaction data. We respectively note that the performance heterogeneity caused by the number of interactions (i.e., data imbalance) is denoted as the natural heterogeneity, which is not the primary focus of this work as illustrated in the second paragraph of the Introduction. Such natural heterogeneity can be approximately measured when we split users into groups where we independently train and evaluate one model per group, i.e., group-wise training (cf. Figure 1). However, we observe that when we train one model on all users' interaction data (i.e., unified training), active users receive significantly more performance gains compared to inactive users. In other words, model training biases might further amplify the natural performance heterogeneity. The ultimate goal of this work is not to achieve similar recommendation quality for active and inactive users, but to alleviate training bias with causal techniques and improve the performance gains of tail users during unified training. We have added more clarifications of the paper motivation in the Introduction.

}

\todo{[DONE] +* 

7. There are some basic writing problems. 
[DONE] For example, in Figure 3, “Observsed” is a typo. 
[DONE] In line 29-31 of page 2, the authors referred to the same paper in one sentence. 
[DONE] Another suggestion is to replace “the above analysis and approximation” before Eq.11 with detailed equations.

}

\todo{[DONE]+*

8. Some recent related works on causal recommendation are missing.

[R1] Yue He, Zimu Wang, Peng Cui, Hao Zou, Yafeng Zhang, Qiang Cui, Yong Jiang:
CausPref: Causal Preference Learning for Out-of-Distribution Recommendation. WWW 2022
[R2] Yichao Wang, Huifeng Guo, Bo Chen, Weiwen Liu, Zhirong Liu, Qi Zhang, Zhicheng He, Hongkun Zheng, Weiwei Yao, Muyu Zhang, Zhenhua Dong, Ruiming Tang:
CausalInt: Causal Inspired Intervention for Multi-Scenario Recommendation. KDD 2022

}

\todo{[DONE]+**

3, I am interesting to dive into why we adopt the way of reducing the performance heterogeneity can enhance the overall performance? Because the head users present more opportunities of buying or visiting items, which would improve the overall performance for the recommender models that pay more attention to the head users and popular items in theory.

We appreciate the reviewer’s rational concern on the clarification of the paper motivation. We agree that improving the recommendation performances for head users is critical for recommender systems and would improve the overall performance. We respectfully note that reducing the performance heterogeneity not necessarily means sacrificing the performances for head users. Our major claim is that the spurious correlations brought by head users would bias the unified training of all users, leading to limited performance gains for long-tail users. As shown in Figure 1, tail users obtain significantly fewer performance gains than head users in unified training compared to group-wise training.
Our primary objective is to address the issue of training bias and enhance the performance of tail users, without compromising the performance of head users. The experimental results depicted in Figure 4 demonstrate that CausalD is a significant step towards achieving this goal. Additionally, we believe that enhancing the recommendation quality for tail users would have a positive impact on the platform ecology in the long term.

}

%
%
%
%

%
%
%


\todo{[DONE]+*

7. The right subfigure in Figure 1 and subfigures in Figure 4(b) have different value scales for behavior consistency.

Thanks for your suggestion. We have corrected the value scales to be consistent.

}

\todo{[DONE]+*

Here some Typo/ Clarifications:
[DONE] 1) Fig.1 title  Inconsisitent-> Inconsistent, Consisitent-> Consistent
[DONE] 2) Page 1 Line 59  heterogenity-> heterogeneity
[DONE] 3) Fig.2 Recommedantion->Recommendation
[DONE] 4) Fig.3 Observsed-> Observed
[DONE] 5) Page 5 Line 5-6 ‘architectures is acceptable for”, Replacing is by are.
[DONE] 6) Tab.1 Line 30 controll->control
    It is better to proofread the paper again.

需要重新用grammarly扫一遍，这个审稿人指出的问题多半是 figure、脚注 中的

}

\begin{figure}[!t] \begin{center}
    \includegraphics[width=\columnwidth]{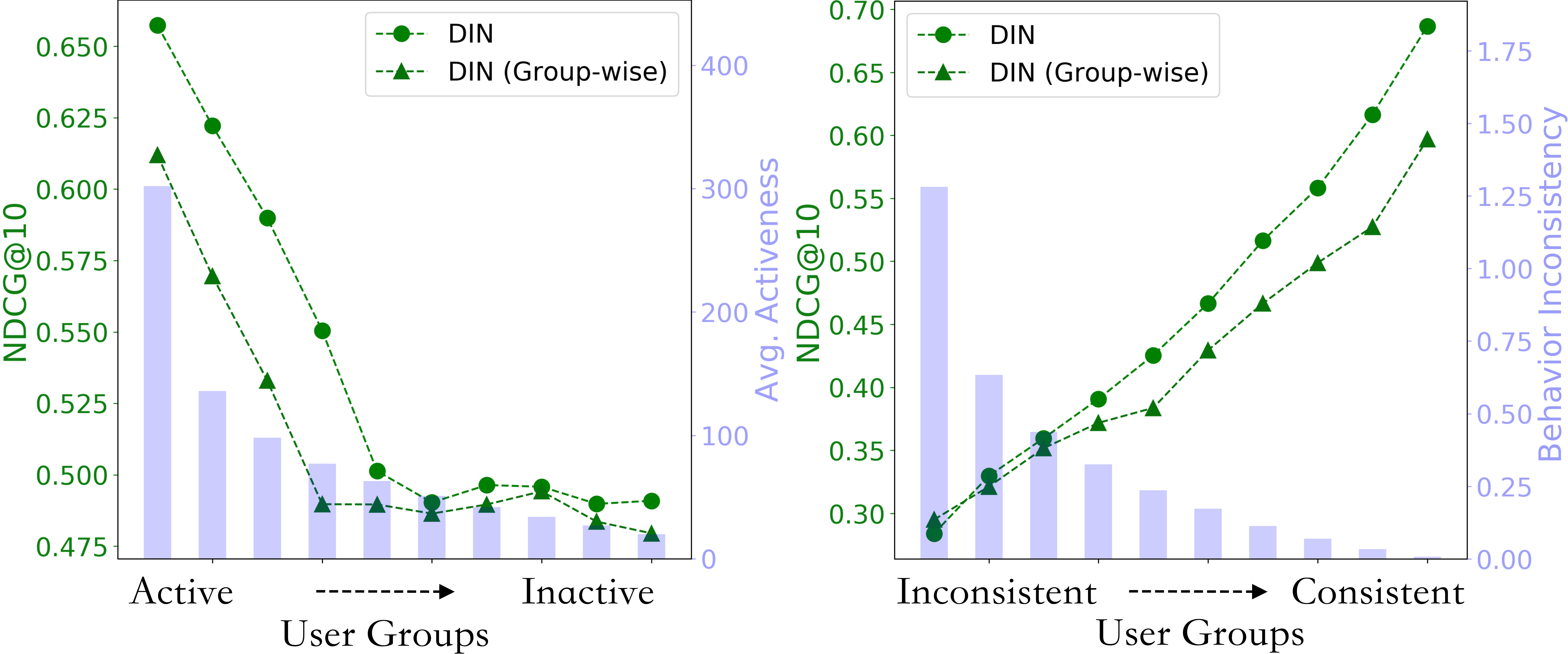}
    \caption{
    	Recommendation performance over user groups clustered by  1) user activeness, and 2) behavior consistency on popular items. DIN~\cite{Zhou_Zhu_Song_Fan_Zhu_Ma_Yan_Jin_Li_Gai_2018} and DIN (Group-wise) are under unified training and group-wise training, respectively.
    	The user number of each group remains the same.
	}
\label{fig:longtail_base}
\end{center} \end{figure}

Recommender systems alleviate the information overload on the Web by providing personalized services for users seeking information.
It has become a default choice to learn recommender models from user historical behaviors~\cite{He_Liao_Zhang_Nie_Hu_Chua_2017,DBLP:journals/tkde/YinWZLZ22,DBLP:journals/tkde/GaoLLLZLJ22,DBLP:journals/tkde/UngerT22}. The model typically exhibits performance heterogeneity with significant divergence across different user groups.
Figure~\ref{fig:longtail_base} provides empirical evidence where 
we train a representative recommendation model DIN~\cite{Zhou_Zhu_Song_Fan_Zhu_Ma_Yan_Jin_Li_Gai_2018} on the benchmark Amazon Review dataset~\cite{McAuley_Targett_Shi_Hengel_2015} and evaluate the performance over equally sized user groups. User groups can be divided according to diverse factors, and we take two factors for illustration: 1) the activeness of users (\ie, the number of interactions during training); and 2) the user behavior consistency on popular items\footnote{We cluster items into eighteen groups by popularity, equally split the behavior sequence of each user into two parts (past and now), and represent each part with item distribution over the clusters. We measure the behavior consistency on popular items of a user with the KL divergence of these two distributions. Intuitively, recommender models will perform better on users with consistent behaviors. \label{fn:behavior}}. In both cases, it is observed that a small portion of head users receive more accurate recommendation results than the others. Moreover, head users enjoy significantly more performance gains under unified training (one model for all groups) than group-wise training (one model for each group). This means that the unified training might aggravate the \textbf{performance heterogeneity} problem among users due to model biases.
In this work, we analyze and address the performance heterogeneity issue in recommender systems, which has thus far received little scrutiny.

There are indeed two sources of the performance heterogeneity issue from the data and model perspectives. A \textit{natural source} is the imbalance of training data distribution over users. For instance, active users are likely to receive relatively accurate recommendations owing to rich interaction records and more comprehensive interest modeling. A \textit{model source} is that recommender models further amplify the impact of data imbalance during training~\cite{Steck_2018}, \ie, the model bias towards the head users.
Accordingly, we set two targets in addressing this performance heterogeneity issue: (1) we want to preserve the natural heterogeneity since forcibly removing it might hurt the overall performance~\cite{Wang_Wang_Beutel_Prost_Chen_Chi_2021}; and (2) we would like to eliminate the model bias that amplifies the heterogeneity. However, most existing methods either neglect or eliminate the heterogeneity, such as the fairness recommendation~\cite{Li_Chen_Fu_Ge_Zhang_2021} that pursues equal performance across different groups\footnote{Note that recommendation methods enforcing fairness also requires known sensitive features, which might be unavailable in the performance \zsy{heterogeneity} problem.}.

\begin{figure}[!t] \begin{center}
    \includegraphics[width=\columnwidth]{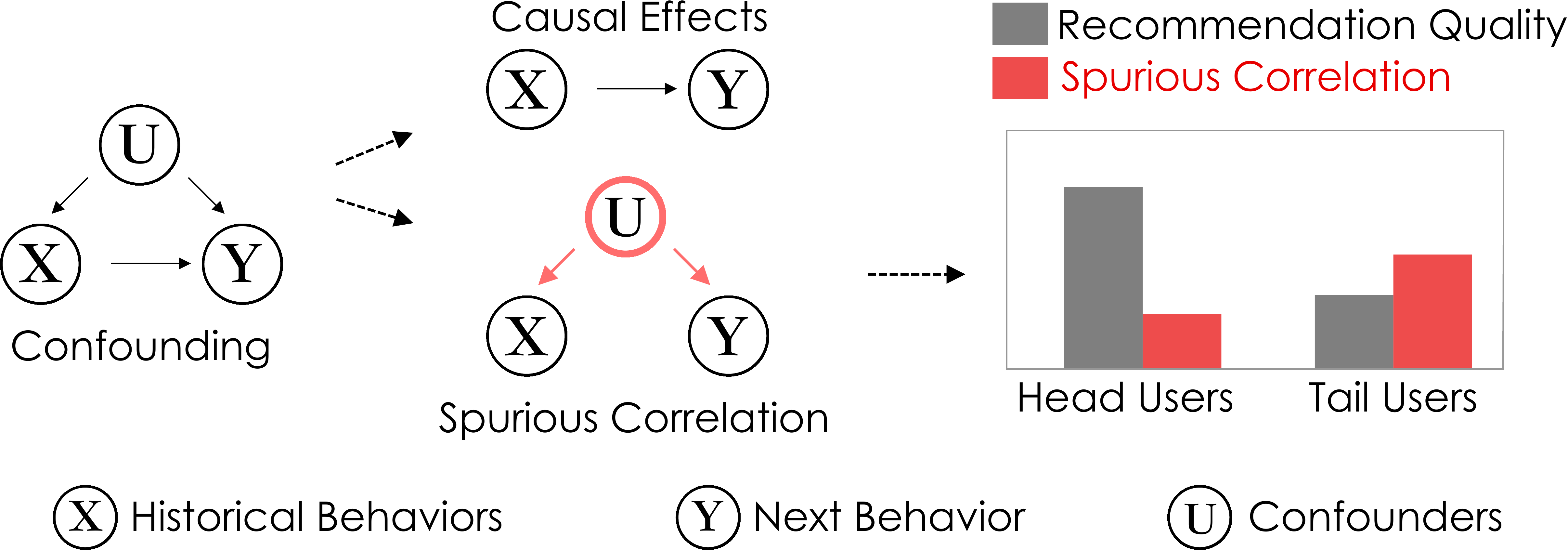}
    \caption{
    	Causal graph for illustrating the amplified performance heterogeneity due to spurious correlation.
	}
\label{fig:causalGraphBias}
\end{center} \end{figure}

To achieve the targets, we dive into the generation process of interactions to uncover the causes of the model bias, which is abstracted as a \textit{causal graph}~\cite{glymour2016causal} as depicted in Figure \ref{fig:causalGraphBias}:
\begin{itemize}[leftmargin=*]
	\item $X \rightarrow Y$. Recommendation models predict user next behavior $Y$ (\eg, clicking an item), based on user historical behaviors $X$ under the assumption that historical behaviors $X$ are the cause of next behavior $Y$ ~\cite{He_Liao_Zhang_Nie_Hu_Chua_2017}. 
	\item $U \rightarrow X, U \rightarrow Y$. $U$ represents a set of factors (\eg, user activeness, and user preference on popular items) that directly influence user historical behaviors and next behaviors apart from the effect $X \rightarrow Y$. For instance, user activeness will increase the size of historical records $U \rightarrow X$ and encourage item exploration $U \rightarrow Y$. User preference on popular items will influence the distribution of user historical behaviors $U \rightarrow X$ and the property of the next item $U \rightarrow Y$. 
\end{itemize}
$U$ is identified as a confounder between $X$ and $Y$, which will lead to spurious correlation $X \leftarrow U \rightarrow Y$ when recommendation models (\eg, sequential recommendation~\cite{DBLP:journals/tkde/ZhengWXLW22,DBLP:journals/tkde/CuiWLZW20} and graph-based recommendation~\cite{DBLP:journals/tkde/FanMLWCTY22}) estimate the correlation between historical behaviors and next behaviors~\cite{glymour2016causal}. The spurious correlation is the source of model bias that amplifies the performance heterogeneity. 
For instance, preferring popular items $U=1$ will lead to next behaviors $Y$ linked with popular items that are not caused by historical behaviors $X$. Such $<$ historical behaviors, next behavior $>$ pairs are spuriously correlated and bias the model to recommend popular items regardless of historical behaviors. Note that such model bias might not do harm to users that prefer popular items (head users). However, other users might suffer from these spurious correlations under the unified training\footnote{The spurious correlation scores in Figure \ref{fig:causalGraphBias} are correlated with the results in Figure \ref{fig:longtail_base} where the experiments are conducted in the Amazon Book dataset. Tail users obtain few gains in unified training due to suffering more spurious correlations.} and receive low-quality recommendations (\cf Figure \ref{fig:causalGraphBias}). As such, the performance heterogeneity issue is amplified due to the confounding.

In this light, we focus on alleviating performance heterogeneity by blocking the spurious correlation $X \leftarrow U \rightarrow Y$, \ie~modeling the causal effect $X \rightarrow Y$. Intuitively, we can achieve the target by applying the debiasing techniques in recommendation that handles confounders such as inverse propensity scoring (IPS) ~\cite{Wang_Liang_Charlin_Blei_2018,Agarwal_Takatsu_Zaitsev_Joachims_2019}, back-door adjustment~\cite{Wang_Feng_He_Wang_Chua_2021,Zhang_Feng_He_Wei_Song_Ling_Zhang_2021}, counterfactual inference~\cite{Wang_Feng_He_Zhang_Chua_2021,Liu_Cheng_Dong_He_Pan_Ming_2020}. To obtain accurate estimation, these methods require either unbiased data or \textit{observations of confounders}, which might not be easily satisfied in real-world recommender systems. For example, collecting unbiased data from random trials (\eg~uniform exposure) is costly due to hurting user experience. Meanwhile, \textit{unobserved} confounders are common in recommendation, breaking the unconfoundedness assumption~\cite{Rubin_1978}. \zsy{For instance, social relationships (\eg, friends and family members) are known to have significant impact on recommender systems, \ie, bringing exposure to items ($U \rightarrow Y$) when friends directly recommend items, and conformity to the users ($U \rightarrow X$). However, whether a click happens due to the user's inherent interests or the suggestions/recommendations from the user's friends or family members is not routinely available for many recommender systems. Another example of unobserved confounder is user's current mood, which would TODO }

%
Hence, there is an urgent need for a debiasing method to deal with unobserved confounders without costly unbiased data.

\todo{[DONE][Semi-DONE]++*

3. The authors used Figure 2 to show that head users and tail users are recommended differently. Can the histogram in Figure 2 be supported by any real dataset? Please introduce the data source of the histogram in Figure 2. Also, Figure 2 analyzes the confounding effects with popularity, but popularity actually is an observed confounder.

Yes, the histograms in Figure 2 are supported by the Amazon Book dataset. The histograms in Figure 2 exhibit an abstraction of Figure 1, where the experiments are conducted on the Amazon Book dataset. We have mainly two observations in Figure 2:
1) Head users enjoy better recommendation quality than tail users. This observation is depicted in Figure 2, which can be intuitive.
2) Head users obtain significantly more performance gains in unified training compared group-wise training than tail users. We attribute this phenomenon to that tail users suffer more from the training bias and the model learns more spuriously correlated user-item pairs for tail users, leading to few gains during unified training.

We acknowledge that item popularity is an observed confounder and we use it as an illustrative example to better explain the concept. In Figure 2, we present histograms that depict an abstraction of Figure 1. In the left figure of Figure 1, the histograms show behavior consistency, which is an indicator of the extent to which users' intentions change and can be an unobserved confounder that is not directly captured by recommender systems.

For better clarity, we have added more illustrations on Figure 2 and the connections between Figure 2 and Figure 1 in the Introduction.

}

\begin{sloppypar}
Towards the goal, we propose to estimate the causal effect $P(Y \mid do(X))$ in recommendation through the front-door adjustment~\cite{glymour2016causal} (FDA) according to the latent representation of user historical behaviors, which is a mediator between $X$ and $Y$ (see Figure~\ref{fig:causalGraph}). 
The core idea of FDA is to estimate the outcome of the intervention $do(X)$ by adjusting the distribution $P(Y \mid do(M))$ over $P(M \mid X)$.
FDA thus takes a two-step estimation procedure: 1) sampling $M=m$ from $P(M \mid X)$; and 2) estimating the causal effect from $M$ to $Y$, \ie~$P(Y \mid do(M=m))$. 
However, these estimations are non-trivial in recommendation for two reasons:  
1) the deterministic encoder in many existing recommendation models can only generate one latent user representation $m$, failing to model the conditional distribution $P(M \mid X)$ and sample multiple $m$; 
and 2) under the strict latency restrictions in recommender systems, the computation cost of enumerating $M$ and $X$ is unaffordable during model inference.
To tackle these two challenges, we encapsulate multi-teacher distillation into the instantiation of FDA, and propose the causal distillation framework, named \textbf{\textit{CausalD}}. CausalD employs the pre-trained heterogeneous models to perform sampling of $M=m$ give $X$ (Challenge 1). Upon the model-based sampling, CausalD estimates the causal effects based on user sampling and multi-teacher ensemble. The estimated causal effects serve as the distillation label for training an efficient student recommendation model, which directly estimates the effect without modeling $P(M \mid X)$ and $(Y \mid do(M=m))$ per $m$ (Challenge 2).
In summary, this work makes the following contributions:
\begin{itemize}[leftmargin=*]
	\item We analyze the performance heterogeneity issue of recommender models from a causal view and address the issue by modeling causal effect with front-door adjustment to handle the \textit{unobserved confounders}.
	\item We propose a \textit{causal multi-teacher distillation} (CausalD) framework, which realizes FDA to estimate the causal effect and preserves the inference efficiency.
	\item We instantiate CausalD on DIN and DeepFM, and conduct extensive experiments on three real-world datasets, validating the effectiveness of our analysis and CausalD.
\end{itemize}
\end{sloppypar}

\section{Preliminaries}



Confounding is the distortion of the association between two variables $X$ and $ Y$, which share a common cause, namely the confounder $U$. The confounder will bring the spurious correlation between $X$ and $Y$ ($X \leftarrow U \rightarrow Y$), leading to the gap between $P(Y \mid X)$ and the causal effect of $X$ on $Y$.
To accurately estimate the causal effect, causal inference performs interventions by introducing the $do$-operator and estimates $P(Y \mid do(X))$ instead of $P(Y \mid X)$. There are mainly two deconfounding techniques that estimate $P(Y \mid do(X))$~\cite{glymour2016causal} from observed data, \ie, back-door adjustment, and front-door adjustment. In this section, we review the main idea of these techniques:

\subsection{Back-door Adjustment} \label{sec:bdasec} Back-door adjustment can handle \underline{observed} confounders $Z$. Note that $U$ represents all confounders. Under Bayes Rule, the posterior $P(Y|X)$ can be written as:
\begin{align}
	P(Y \mid X) = \sum_z P(Y \mid X, Z = z) \underline{P(Z = z \mid X)}.
\end{align}
In this modeling, there is a back-door path between $X$ and $Y$, which refers to the indirect path from $X$ to $Y$ that contains an arrow to $X$, \ie, $X \leftarrow Z \rightarrow Y$. To block the back-door path, back-door adjustment adjusts the parent variables of $X$ to make these variables independent of $X$ with the $do$-operator. As such, $X \leftarrow Z$ is cut due to the independence, \ie, forcibly assigning a target value to $X$ regardless of its parent $Z$. Formally,
\begin{equation}
\begin{aligned}
	P(Y \mid do(X)) &= \sum_z P(Y \mid do(X, Z = z)) P(Z = z \mid do(X)),   \\
	                &= \sum_z P(Y \mid X, Z = z) \underline{P(Z = z)}.\label{eq:bd_original}
\end{aligned}
\end{equation}
The conditional probability $P(Y \mid X, Z = z)$ is invariant to adjustment since how $Y$ responds to $X$ and $Z$ is regardless of whether the change of $X$ is independent of $Z$. As such, $P(Y \mid do(X, Z = z)) = P(Y \mid X, Z = z)$. 
In essence, back-door adjustment computes the association between $X$ and $Y$ for each value $z$ of $Z$, and takes the average over the prior probability $P(Z = z)$. The transition from $P(Z = z \mid X)$ to $P(Z = z)$ indicates that the change of $X$ no longer affects the distribution of choosing $Z$, thus obeying the randomized experiments where we can accurately estimate the causal effect of $X$ on $Y$.

\begin{figure}[!t] \begin{center}
    \includegraphics[width=\columnwidth]{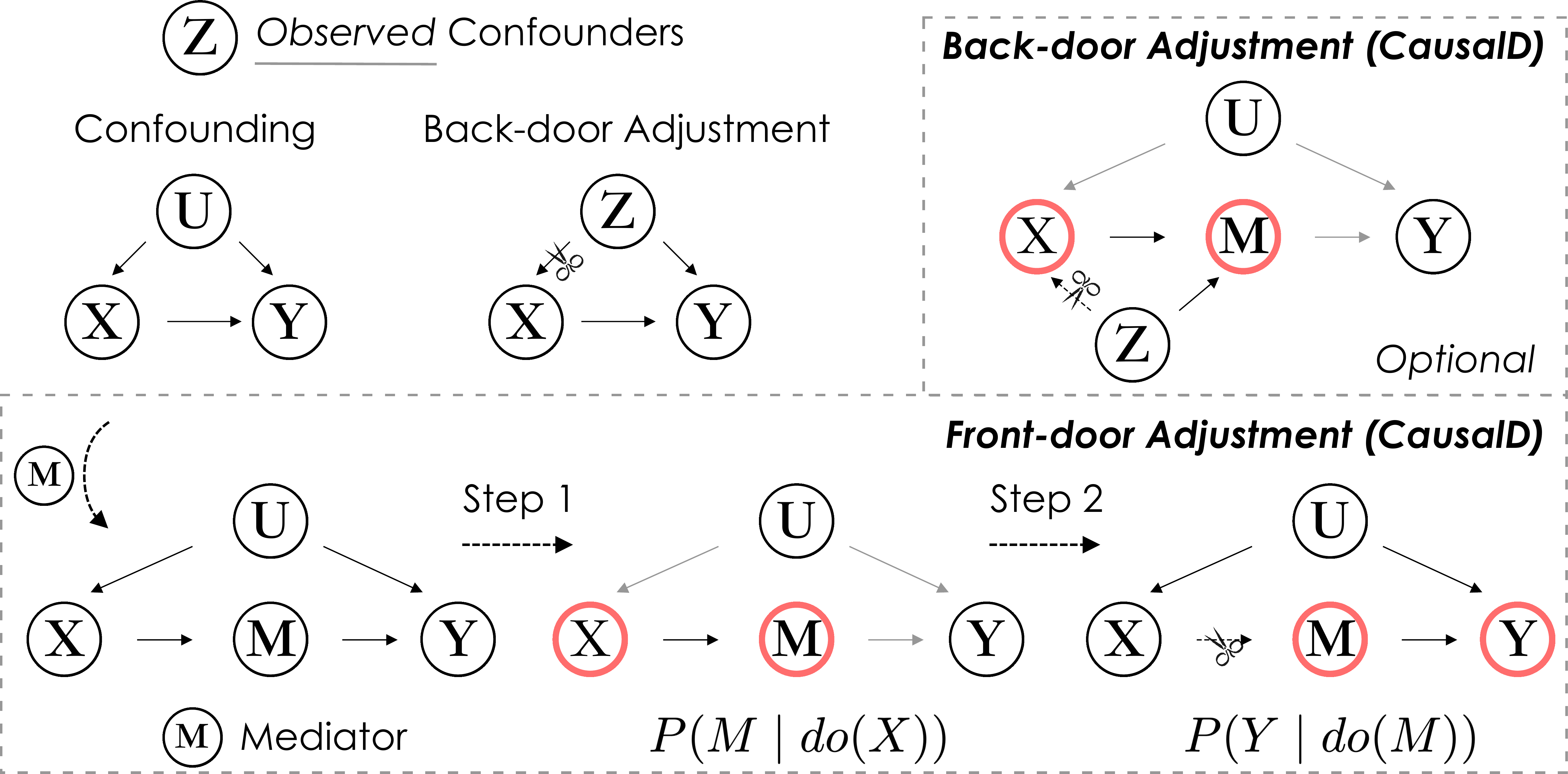}
    \caption{
    	Causal graph for illustrating the back-door adjustment and front-door adjustment.
	}
\label{fig:causalGraph}
\end{center} \end{figure}

\subsection{Front-door Adjustment} \label{sec:fdasecsec}

From Equation (\ref{eq:bd_original}), we can find that the back-door adjustment requires the knowledge of the confounder variable $Z$, \ie, the detailed values $z$. However, there are many unobserved confounders in $U$ that are hard to measure in practice, such as the \zsy{social relationships that might bring exposure to items (\eg, friends recommending items) and conformity to users,} in recommender systems. Fortunately, front-door adjustment can estimate the causal effects by introducing a proper mediator variable $M$ between $X$ and $Y$. By Bayes Rule, with the mediator, the posterior with $do$-operator can be written as:
\begin{align}
	 & P(Y \mid do(X))  = \sum_{m} P(M = m \mid do(X)) P(Y \mid do(M = m)). \label{eq:bayesM}
\end{align}
Then, FDA can be decomposed into the following two steps:
\begin{enumerate}[leftmargin=*]
	\item Estimating $P(M = m \mid do(X)) $. $P(M = m \mid do(X)) = P(M = m \mid X)$ since the back-door path $X \leftarrow U \rightarrow Y \leftarrow M$ is naturally blocked thanks to the colliding effect $U \rightarrow Y \leftarrow M$. An uncontrolled collider variable $Y$ will block the effect from $U$ to $M$~\cite{glymour2016causal}. 
	\item Estimating $P(Y \mid do(M = m))$ \wrt the unblocked back-door path $M \leftarrow X \leftarrow U \rightarrow Y$. Similar to the back-door adjustment, the remedy is to adjust the parents of $M$, \ie, $X$ and make $(X,M)$ independent so that we can block the back-door path by cutting $M \leftarrow X$.
\end{enumerate}
By the Bayes Rule, the original likelihood can be written as:
\begin{equation}
\begin{aligned}
	 & P(Y \mid X)  = \\
	 & \sum_{m} P(M = m | X) \sum_{x} \underline{P(X = x | M = m)}[P(Y | M = m, X = x)].
\end{aligned}
\end{equation}
By adjusting $X$, FDA estimates the causal effects as follows:
\begin{equation}
\begin{aligned}
	 & P(Y \mid do(X)) = \\
	 & \sum_{m} P(M = m \mid X) \sum_{x} \underline{P(X = x)} [P(Y \mid M = m, X = x)]. \label{eq:original_fd}
\end{aligned}
\end{equation}
For brevity, we omit the transition details which share similar spirits to Equation (\ref{eq:bd_original}). In essence, after adjustment, $M$ becomes independent of $X$ and $P(X = x \mid do(M = m)) = P(X = x)$, thus blocking the path $ Y \leftarrow U \rightarrow X \rightarrow M$. More details of FDA can be found in \cite{glymour2016causal}. To use FDA, a proper $M$ should satisfy the following three requirements:

\begin{enumerate}[label=(\roman*)]
	\item $M$ intercepts all directed paths from $X$ to $Y$; \label{enu:FD1}
	\item there is no unblocked back-door path from $X$ to $M$; \label{enu:FD2}
	\item all back-door paths from $M$ to $Y$ are blocked by $X$. \label{enu:FD3}
\end{enumerate}

\section{Method}

\todo{[DONE][Semi-DONE]+*

3. This work mainly focuses on alleviating the performance heterogeneity problem. The formal definition of this particular problem is suggested to introduce in Section 3.1 Problem Formulation.

Thanks for your suggestion. We have added the problem formulation of performance heterogeneity alleviation in Section 3.1.


}

\todo{[DONE] +++***

1, The authors claimed that the latent factors derived from the historical behaviors X can be regarded as mediator M from the respective of causal graph, which could be argued. It is reasonable to obtain the latent factor for predicting user’s preferences by utilizing an advanced encoder like MLP. However, it is not a convincing to state that the encoder can extract the mediator without some additional assumptions or theoretical analysis.

We acknowledge that the relationship between the latent factors derived from historical behaviors and the mediator M is not necessarily straightforward, and that additional assumptions or theoretical analysis may be required to fully understand the nature of this relationship.

However, we would like to clarify that our claim is based on the assumption that the latent factors represent a set of features that are relevant to the underlying causal mechanism linking historical behaviors to user preferences. This assumption is grounded in the notion that historical behaviors are indicative of user preferences, and that the latent factors capture the underlying patterns and structures in the historical behaviors that give rise to these preferences.

In terms of our methodology, we use advanced neural network models such as MLP for mediator extraction in order to build an end-to-end system that is easy to train and deploy. While we understand the importance of theoretical guarantees, we believe that our approach is empirically effective in identifying relevant features that mediate the relationship between historical behaviors and user preferences.

That being said, we agree that additional theoretical analysis is an important area for future work. We appreciate your feedback and will consider your concerns in our future research.

}

\todo{[DONE]+**

2, Another important concert is that the mediator M can intercept all directed paths from
X to Y. The issue is analogous to above concert in casual graph. It is benefit for authors to provide an intuitive or theoretical explanation of why we expect this to happen would be good.

Thanks for your suggestion. We have accordingly  added the following illustrations in Section 3.2:

Intuitively, if this requirement is not satisfied, there are other direct or indirect effect of $X$ on $Y$ besides $X \rightarrow M \rightarrow Y$, \eg, $X \rightarrow M^* \rightarrow Y$ or $X \rightarrow Y$. Then, front-door adjustment over each path (\eg, cutting $X \rightarrow M$ for $X \rightarrow M \rightarrow Y$ in step 2) would leave other paths $X \rightarrow M^* \rightarrow Y$ or $X \rightarrow Y$ still confounded by $U$, thus failing to estimate the causal effects.

}

\subsection{Problem Formulation} \label{sec:problemformu}

\vpara{Recommendation.} A recommendation dataset $\mathcal{D} $ contains user-item interaction records $\{ (x_i, y_i, o_i) \}_{ i = 1, \dots, |\mathcal{D}|}$ logged in a recommender system. $x_i$ denotes the user data such as user ID and user historical behaviors. $y_i$ denotes the item data such as item ID. $o_i \in \{ 0,1 \}$ is a binary factor indicating whether the user has interacted with the item. Given user data $x$ and item data $y$, a recommendation model is learned to predict how likely the user will interact with the item, and the prediction score is used for candidate item ranking. Top-ranked items will appear in the recommendation list and are sorted according to the ranking scores.

\vpara{\zsy{Performance Heterogeneity.}} \zsy{We define the total performance heterogeneity as the standard deviation of recommendation performances on different user groups' datasets $\{ \mathcal{D}_e  \}_{e \in \mathcal{E}}$ under \textbf{\textit{unified training}}, \ie, }
\begin{align}
\color{myblue}	S_h = \sqrt{\frac{1}{|\mathcal{E}|-1} \sum_{e \in \mathcal{E}} ( S(\mathcal{D}_e)- {\bar S} )^2},
\end{align}
\zsy{where $S(\mathcal{D}^e)$ denotes the recommendation performance on the $\mathcal{D}^e$ dataset and ${\bar S}$ denotes the average performance on all datasets. Such performance heterogeneity includes the natural heterogeneity due to data imbalance across user groups, and the model bias that amplifies the natural heterogeneity and hurts the performance of unprivileged user groups. We define the natural performance heterogeneity as the standard deviation of recommendation performances on different user groups' datasets under \textbf{\textit{group-wise training}}. Specifically, we independently train and evaluate one model per group where the performance is denoted as $S^* (\mathcal{D}_e)$. The natural performance heterogeneity can be written as: }
\begin{align}
\color{myblue}	S_h^* = \sqrt{\frac{1}{|\mathcal{E}|-1} \sum_{e \in \mathcal{E}} ( S^*(\mathcal{D}_e)- {\bar S}^* )^2},
\end{align}
\zsy{We define the performance heterogeneity brought by the model bias, \ie, bias-amplified performance heterogeneity as:}
\begin{align}
	\color{myblue} S_h^{\circ} = S_h - S_h^*. \label{eq:training_hetero}
\end{align}

In this paper, we use capital letters ($\eg, X$) to denote variables, lowercase letters ($\eg, x$) to denote particular values of variables, and bold ones ($\eg, \mathbf{x}$) to denote the vectorial representations.

\todo{[DONE]++*

5. The authors did not introduce why distillation is a better choice for front-door adjustment principle than other methods. There is an obvious gap between Challenge 1&2 and why distillation is a good choice for front-door adjustment principle.

We have added the following clarifications in Section 3.3 to bridge the gap between Challenge 1&2 and the knowledge distillation technique:

As for \textbf{challenge 2}, we approximate the computationally expensive expectation calculation (\cf Equation (\ref{eq:original_fd})), which enumerates each sampled $m$ and $x$ with knowledge distillation. That is, we first estimates the causal effects using heterogeneous teacher models (computationally expensive), and transfer the knowledge into an efficient student model (computationally light-weight). The student directly estimates the effect while trying to approaching the causal effect estimated by teachers via distillation losses.

}

\subsection{Front-door Adjustment in Recommendation} \label{sec:causalview}

In this subsection, we analyze the critical challenges in satisfying the three requirements of FDA and implementing FDA in deep recommendation models. Recall that in recommendation, $X$ refers to the historical behaviors while $Y$ refers to the next behavior (\textit{c.f.} Figure \ref{fig:causalGraph}). Many recommendation models follow the encoder-predictor structure where the encoder transforms $X$ into latent factors, and the predictor determines the correlations between $X$ and $Y$ based on the latent factors and the target-item feature. As such, it is intuitive to view the latent factors obtained from $X$ as the mediator $M$, and the representation of latent factors as the mediator feature. We analyze whether such $M$ satisfies the requirements one by one:

 \vpara{Requirement (i).}  $M$ should intercept all directed paths from $X$ to $Y$. This means that there is no other information of $X$ that could contribute to the recommendation prediction but is not encoded in $M$. \zsy{Intuitively, if this requirement is not satisfied, there are other direct or indirect effect of $X$ on $Y$ besides $X \rightarrow M \rightarrow Y$, \eg, $X \rightarrow M^* \rightarrow Y$ or $X \rightarrow Y$. Then, front-door adjustment over each path (\eg, cutting $X \rightarrow M$ for $X \rightarrow M \rightarrow Y$ in step 2) would leave other paths $X \rightarrow M^* \rightarrow Y$ or $X \rightarrow Y$ still confounded by $U$, thus failing to estimate the causal effects.} In the encoder-predictor recommendation architecture, the latent factors $M$ \textbf{as a whole} contain all the information of the historical behaviors $X$ which are used to make predictions.
 As such, the latent factors $M$ as a whole obtained from the encoder satisfy the requirement. 



 \vpara{Requirement (ii) \& (iii).} These requirements indicate that there are no confounders affecting $(X,M)$ and no confounders other than $U$ affecting $(M,Y)$, which are mostly satisfied if we use the latent factors obtained by the encoder as the mediator. Specifically, the parameters of the encoder and the predictor are randomly initialized, thus containing no confounder-related information before training. During training, the parameters are learned from $(X,Y)$ pairs from the recommendation datasets, and can only encode confounders that affect $(X,Y)$, which are all included in $U$ of the back-door path $X \leftarrow U \rightarrow Y$. As such, there are no confounders other than $U$ affecting $(M,Y)$. $(X,M)$ are not affected by the back-door path $X \leftarrow U \rightarrow Y \leftarrow M$ thanks to the colliding effect (\textit{c.f.} Section \ref{sec:fdasecsec}). In addition, we show that we can improve the satisfaction of Requirement (ii) with causal feature distillation in Section \ref{sec:featuredistillation}.


With the mediator, Equation (\ref{eq:original_fd}) is an expectation over the conditional distribution $P(m \mid X)$, and the prior distribution $P(x)$. First, we should sample different values $m$ of $M$ given $X$ while a deterministic encoder can only output one $m$ given user data $x$. Therefore, the sampling of $M$ given $X$ remains a challenge (\textbf{Challenge 1}).
Second, the sample spaces of these two distributions are infinite, which makes the expectation calculation intractable. Moreover, the calculation of $P(Y \mid m,x)$ for each $m$ and $x$ is computationally expensive.
Therefore, how to derive a proper approximation for expectation calculation, while preserving the \textbf{inference efficiency} due to the latency restrictions in recommender systems, presents the second challenge (\textbf{Challenge 2}).

\subsection{Causal Multi-teacher Distillation} 

The ultimate goal of our method is to learn a recommendation model that could efficiently estimate the causal effect of user historical behaviors on the next behavior by handling unobserved confounders with FDA. The implementation of FDA faces the above two challenges in recommendation. To address \textbf{Challenge 1}, we propose to incorporate heterogeneous recommendation models where each model yields one mediator value $m$. 
\zsy{As for \textbf{challenge 2}, we approximate the computationally expensive expectation calculation (\cf Equation (\ref{eq:original_fd})), which enumerates each sampled $m$ and $x$ with knowledge distillation. That is, we first estimates the causal effects using heterogeneous teacher models (computationally expensive), and transfer the knowledge into an efficient student model (computationally light-weight). The student directly estimates the effect while trying to approaching the causal effect estimated by teachers via distillation losses.}
By unifying heterogeneous teachers and knowledge distillation, we propose the causal multi-teacher distillation framework, namely \textbf{\textit{CausalD}}. 
In the following, we dive into the technical details of constructing heterogeneous teachers, performing FDA with heterogeneous teachers, and how we define the distillation losses. Key notations are summarized in Table \ref{tab:notations}.

%
%

\subsubsection{Constructing Heterogeneous Teachers}
We can split training recommendation data samples $\mathcal{D}$ into non-IID subsets $\{ \mathcal{D}_1, \mathcal{D}_2, \dots, \mathcal{D}_{ K }  \}$ via biased selection \wrt some user attributes or unsupervised clustering. For example, we can select data samples linked with male and female users into two subsets, which have different data distributions due to the interest discrepancy between male and female users. In an extreme case, the user attribute is the user ID where each subset corresponds to a particular user. 
In practice, we choose the attribute related to an observed confounder of interest, such as user activeness or user behavior consistency. We illustrate that such a split could improve feature distillation in Section \ref{sec:featuredistillation}.
We train one teacher model per subset, resulting in $K$ heterogeneous teacher models that encode different data information.
Each teacher contains an encoder, which extracts latent factors (mediator) from user historical behaviors, and a predictor, which makes recommendation predictions based on the latent factors and the target-item feature. We denote the encoders of pre-trained heterogeneous teacher models as $\Phi=\{\phi_k\}_{k=1, \dots, K}$, which will be used for obtaining mediator values. In the experiments (\cf Section \ref{sec:indepthAnalysis}), we show that the CausalD framework significantly benefits more from heterogeneous teachers than homogeneous teachers trained on IID subsets, validating the rationality of constructing heterogeneous teachers for sampling mediator values. 

The model architecture of each teacher remains the same as the student model. 
Therefore, the models are \textit{\textbf{unprivileged}} teachers \wrt the student. A teacher is unprivileged if it has the same model architecture as the student and is trained on a subset of the dataset used to train the student.
%
In other words, unprivileged teachers will bring no additional dataset knowledge nor model structure knowledge during distillation. This design permits a fair comparison with baseline models trained without distillation. In practice, incorporating additional features or higher-capacity model architectures \zsy{are} acceptable for FDA and might further improve the recommendation performance, which is beyond the scope of this paper.

\subsubsection{Front-door Adjustment for Causal Label Distillation} \label{sec:fdaCLD}


To estimate the causal effect with FDA, we sample multiple mediator values $m$ with heterogeneous teachers and multiple $x$ using the in-batch sampling strategy. We then calculate the causal effect as the expectation of $P(Y \mid M=m, X=x)$, which is modeled as a neural network, over sampled $m$ and $x$ (\textit{c.f.} Equation (\ref{eq:original_fd})). The causal effect is the distillation label for training the student recommendation model.
Specifically, the expectation calculation in Equation (\ref{eq:original_fd}) can be decomposed into the following sub-processes:

\vpara{Sampling of $M=m$.} We use heterogeneous teacher recommendation models to do the sampling $M$ given $X$. Given input $x_i$, the encoders $\Phi=\{\phi_k\}_{k=1, \dots, K}$ of heterogeneous teacher models yield heterogeneous $M=m$:
\begin{align}
	\mathbf{\tilde m}_{k,i} = \phi_k(x_i), \quad k = 1, \dots, K, \label{eq:mediator_repre}
\end{align}
where $\mathbf{\tilde m}_{k,i}$ is the feature of mediator value ${m}_{k,i}$ encoded by $\phi_k$ given the input $X=x_i$.
 As for the modeling of $P(M=m_{k,i} \mid X)$, we take a parametric solution, \ie, attention mechanism, and view the attention weights of $X$ on multiple $m$ as the conditional probability.
\begin{align}
	\alpha_{k,i} = \frac{\exp( \mathbf{W}_1 \mathbf{\tilde m}_{k,i} \cdot \mathbf{W}_2 \mathbf{\hat m}_{i} )}{ \sum_k \exp( \mathbf{W}_1 \mathbf{\tilde m}_{k,i} \cdot \mathbf{W}_2 \mathbf{\hat m}_{i} ) }, \quad \mathbf{\hat m}_{i} = \phi^*(x_i)	, \label{eq:pmx}
\end{align}
where $\mathbf{\tilde m}_{k,i}$ is the mediator feature for the $i$-th data extracted by the $k$-th teacher model and is computed by Equation (\ref{eq:mediator_repre}). $\mathbf{\hat m}_i$ is the mediator feature of the $i$-th data computed by the encoder of the student model, \ie, $\phi^*$.

\vpara{Sampling of $X=x$.} 
We take the in-batch sampling strategy for efficiency. Specifically, for a given training sample $x_i$, we take other training samples $\{ x_j \}_{j \neq i}$ in the batch as the sampled $X$. For the prior probability of $X=x_j$, we have no access to the data generation process in the real world. Alternatively, since each sample is randomly taken into the batch, we take the uniform distribution for $P(X = x_j)$.

\begin{table}[t]
\centering
\caption{
    Notations
}
\small
\begin{tabular}{c|p{6cm}}
\toprule
\multicolumn{1}{c|}{ \textbf{Notation} }        & \multicolumn{1}{l}{ \textbf{Description} }          \\
\midrule

$X, Y, Z, M$ & variables denoting the historical behaviors, next behavior, confounder, mediator \\ \midrule
$x, y, z, m$ & particular values of $X, Y, Z, M$ \\ \midrule
$\mathbf{x}, \mathbf{y}, \mathbf{z}, \mathbf{m}$ & vectorial representation of $x, y, z, m$ \\ \midrule
$\phi_k, \phi^*$ & encoder of the $k$-th teacher / student recommendation model \\ \midrule
$\mathbf{\tilde m}_{k,i}, \mathbf{\hat m}_{i}$ & mediator feature extracted by $\phi_k$ and   $\phi^*$ that take $x_i$ as input  \\ \midrule
$ \mathbf{\tilde m}^{ \circ }_i $  & mediator feature estimated by teachers with back-door adjustment     \\ \midrule

$o, {\tilde o}, {\hat o}$ &  recommendation label  (\eg, CTR), prediction estimated by teachers with front-door adjustment, prediction made by the student  \\ \midrule
$ \psi_{FDA} $ & fully-connected layers for estimating $ {\tilde o} $ with front-door adjustment \\ \midrule
$ \psi^* $ & fully-connected layers for estimating $ {\hat o} $  \\ \midrule
$ \mathbf{W}_1, \mathbf{W}_2 $ & trainable matrices for estimating the conditional probability $P(M=m \mid X)$ \\ \midrule
$ \mathcal{L}_{BDA} $ & \textbf{feature} distillation loss with back-door adjustment \\ \midrule
$ \mathcal{L}_{FDA} $ & \textbf{label} distillation loss with front-door adjustment \\ \midrule
$ \mathcal{L}_{Rec} $ & original recommendation loss (\eg, BPR loss, Sampled Softmax loss) \\ \midrule
$ \lambda_{BDA}, \lambda_{FDA} $ &  coefficients to \zsy{control} $\mathcal{L}_{BDA}$, $\mathcal{L}_{FDA}$   \\

\bottomrule
\end{tabular}
\label{tab:notations}
\end{table}

\vpara{Modeling of $P(Y \mid M=m_{k,i}, X=x_j)$.} In recommender systems, the final prediction is generally pair-wise matching. For a user input $x_i$ and a target item $Y=y_i$, the recommendation prediction is mostly a binary classification where the output ${ \hat o }_i$ indicates whether the user will interact with the item (\eg, click or purchase).
Therefore, we can parameterize $P(Y=y \mid M=m_{k,i}, X=x_j)$ as a neural network $\varphi(\cdot)$ followed by a sigmoid layer $\sigma$:
\begin{align}
	P(Y=y_i \mid X=x_j,M=m_{k,i}) = \sigma(\varphi(x_j, \mathbf{\tilde m}_{k, i}, \mathbf{y}_i)), \label{eq:pyxm}
\end{align}
where $\mathbf{\tilde m}_{k,i}$ and $\textbf{y}_i$ are the vectorial representations of $m_{k,i}$ and $y_i$. $\mathbf{\tilde m}_{k,i}$ is extracted by the encoder of the $k$-th teacher recommendation model. 
In recommendation models, $\mathbf{\tilde m}_{k,i}$ can be the user representation extracted from the $i$-th user input data $x_i$ (\eg, user historical behavior sequence), and 
$\mathbf{y}_i$ can be the embedding of item $y_i$ obtained from a trainable embedding look-up table.
Since $M$ intercepts all the direct effects from $X \rightarrow Y$, we can use the mediator $m_j$ to replace $x_j$ and use the vectorial representation $\mathbf{\tilde m}_j$ in the modeling:
\begin{align}
	P(Y=y_i \mid X=x_j,M=m_{k,i}) = \sigma(\varphi(\mathbf{\tilde m}_j, \mathbf{\tilde m}_{k,i}, \mathbf{y}_i)),
\end{align}
where $\mathbf{\tilde m}_j = 1/K \sum_k \mathbf{\tilde m}_{k,j} $ is the mediator feature of data $x_j$ by taking the average of the mediator features of all teachers.
Function $\varphi$ can be modeled by any neural network architecture that makes recommendation predictions. Taking DIN \cite{Zhou_Zhu_Song_Fan_Zhu_Ma_Yan_Jin_Li_Gai_2018} as an example, DIN uses fully-connected layers (FCL) to make click-through-rate predictions with concatenated user-item representations as input. Following this work, we use a FCL network $\psi_{FDA}$ to model $\varphi$:
\begin{align}
	\varphi(\mathbf{\tilde m}_j, \mathbf{\tilde m}, \mathbf{y}) = \psi_{FDA} \left( \left[\mathbf{\tilde m}_j, \mathbf{\tilde m}_{k,i}, \mathbf{y}_i \right] \right), \label{eq:avgmediator1}
\end{align}
where $[\cdot]$ denotes vector concatenation operation. 

\vpara{Estimating $P(Y|do(X))$.} \zsy{With the sampling of $M = m$ (\cf Equation (\ref{eq:mediator_repre}) - (\ref{eq:pmx})), the sampling of $X = x$, the modeling of $P(Y \mid M=m_{k,i}, X=x_j)$ (\cf Equation (\ref{eq:pyxm}) - Equation (\ref{eq:avgmediator1}))}, we can estimate the causal effect of user data $x_i$ on the next item $y_i$ \zsy{by implementing FDA (\cf Equation (\ref{eq:original_fd}))} as follows:
\begin{gather}
	{\tilde o}_{i} = \sum_k \alpha_{k,i} \sum_j \frac{1}{N_b} \sigma( \psi_{FDA} \left( \left[\mathbf{\tilde m}_j, \mathbf{\tilde m}_{k,i}, \mathbf{y}_i \right] \right) ), \label{eq:fd_pred}
\end{gather}
where $\alpha_{k,i}$ is computed with Equation (\ref{eq:pmx}), and ${\tilde o}_{i}$ is the causal recommendation prediction indicating the probability of the interaction between $x_i$ and $y_i$. 

\vpara{Causal Label Distillation.}  ${\tilde o}_{i}$ is predicted by all teacher recommendation models, and serves as the \textbf{causal label guidance} for training the student recommendation model. The prediction of the student model is obtained as follows:
\begin{align}
	{\hat o}_{i} = \psi^* ( \left [\mathbf{\hat m}_{i}, \mathbf{y}_i \right ] ), \label{eq:studentpred}
\end{align}
where $\psi^*$ denotes the predictor of the student recommendation model, which is modeled as another FCL network. $\mathbf{\hat m}_{i}$ is the mediator feature extracted by the student encoder $\phi^*$. Since the Equation (\ref{eq:fd_pred}) introduces trainable parameters \ie, $\mathbf{W}_1$, $\mathbf{W}_2$, and $\phi$, we try to pull the prediction closer to the ground-truth label $o_i$.
Then, the loss function of FDA can be:
\begin{align}
	\mathcal{L}_{FDA} = \sum_i (\underbrace{L_{kd}({\tilde o}_i, {\hat o}_i)}_{\text{Distillation}} + \underbrace{L_{ce}({\tilde o}_i, {o}_i)}_{\text{Consistency}}), \label{eq:fd_obj}
\end{align}
where $L_{kd}$ is the knowledge distillation criterion~\cite{DBLP:journals/corr/HintonVD15}. $L_{ce}$ is the cross-entropy loss. It is noteworthy that the consistency loss solely affects $\mathbf{W}_1$, $\mathbf{W}_2$, and $\psi_{FDA}$. The distillation loss is computed when $\mathbf{W}_1$, $\mathbf{W}_2$, and $\psi_{FDA}$ are fixed.

\todo{[DONE]++*

4. The CausalD framework incorporates the causal label distillation powered by front-door adjustment, which is the primary focus, and the causal feature distillation powered by the back-door adjustment, which is claimed to be optional. Is there any underlying connection between these two objectives?

Yes. The underlying connections between causal label distillation and causal feature distillation can be two-fold:
1) Front-door adjustment and back-door adjustment are two representative causal techniques for deconfounding with unobserved and observed confounders, respectively. We instantiate these two techniques in one unified neural model by identifying mediators and observed confounders. We empirically find that the joint modeling contributes to the overall effectiveness.
2) As described in Section 3.2, the requirement (ii) of applying front-door adjustment is that no confounders simultaneously affect $X$ and $M$. We further apply the back-door adjustment for deconfounding the estimation of causal effects of $X$ on $M$ (cf. Figure 3 Back-door Adjustment). 

We have added these clarifications in a subsection in Section 3.5.

}

\subsubsection{Feature Distillation} \label{sec:featuredistillation} Besides label distillation, we also incorporate the feature distillation loss for training the student recommendation model. Note that the mediator values $m$ extracted by teachers are intermediate features, which can serve as the feature guidance. The feature distillation loss can be defined as:
\begin{align}
	\mathcal{L}_{FT} = \sum_i d ( \pi ( \{ \mathbf{\tilde m}_{k,i} \}_{k=1}^K ) , \mathbf{\hat m}_i), 
\end{align}
where $\pi$ denotes the function that aggregates the intermediate features of teachers.  $d$ denotes a distance function such as Euclidean distance. $\pi$ is decided according to whether we can observe the confounder of interest. That is, if the confounder of interest is not observed, we take the vanilla feature distillation. If the confounder of interest is observed, we employ back-door adjustment to improve deconfounding in the feature level, namely causal feature distillation.

\vpara{Vanilla Feature Distillation.} The aggregation function in the vanilla feature distillation takes the average of all teacher features, which can be written as follows:
\begin{align}
	\pi ( \{ \mathbf{\tilde m}_{k,i} \}_{k=1}^K ) = \frac{1}{K} \sum_k \mathbf{\tilde m}_{k,i}.
\end{align}

\vpara{Causal Feature Distillation.} Besides the vanilla feature distillation, we propose an optional causal feature distillation to handle some confounders $Z$ of interest that can be observed in recommender systems with back-door adjustment (\textit{c.f.} Section \ref{sec:bdasec}). First, we can split the data \wrt different confounder values. By taking user activeness as an example, we can count the number of interactions and accordingly split users into several groups. Since data samples in the $k$-th group are selected when the confounder variable is set to a particular value $Z = z_k$, the parameters of the corresponding models learn the knowledge under $Z = z_k$. Then, teachers can be regarded as the estimators $P(M \mid X, Z = z_k)$, and together approximate $P(M \mid do(X))$ as the weighted sum of the estimation of different models according to Equation (\ref{eq:bd_original}). Formally, we have:
\begin{align}
	 \mathbf{\tilde m}^{ \circ }_i = \mathbb{E}_{z_k} \left [ \phi_k(x_i) \right ] =  \sum_k \mathbf{\tilde m}_{k,i} P(z_k). \label{eq:bddo}
\end{align}
We set $P(z_k)$ to be proportional to the number of users of the group where $\phi_k$ is trained. 
We take $\mathbf{\tilde m}_i^{ \circ }$ as the \textbf{causal feature guidance} for training the student as follows:
\begin{align}
	\mathcal{L}_{BDA} = \sum_i d(\mathbf{\tilde m}^{ \circ }_i, \mathbf{\hat m}_i) . \label{eq:bddistill}
\end{align}
This modeling can better satisfy Requirement (ii) in FDA by actively blocking back-door paths from $X$ to $M$, to meet the complex environments in real-world recommender systems. 
We also replace the average mediator feature $\mathbf{\tilde m}_{j}$ in Equation (\ref{eq:avgmediator1}) - (\ref{eq:fd_pred}) with $\mathbf{\tilde m}^{ \circ }_j$ to improve deconfounding since $\mathbf{\tilde m}^{ \circ }_j$ represents the debiased mediator feature.

\subsection{Model Training}

In summary, we first split the recommendation dataset $\mathcal{D}$ into non-IID subsets $ \{ \mathcal{D}_1, \mathcal{D}_2, \dots, \mathcal{D}_{ K }  \} $ with biased section \wrt some user attributes.
If we aim to address the model bias caused by a particular confounder, we can split the data according to the confounder values.
We train one recommendation model per subset as one teacher, and obtain their encoders $\{ \phi_1, \phi_2, \dots, \phi_K \}$. Given $ (x_i, y_i, o_i) $, we first obtain causal feature guidance $\mathbf{ \tilde m }_i^{\circ}$ from all teachers with Equation (\ref{eq:bddo}), and the mediator feature $\mathbf{ \tilde m}_i$ extracted by the student encoder $\phi^*$. As such, the \textbf{causal feature distillation} loss $\mathcal{L}_{BDA}$ is computed with Equation (\ref{eq:bddistill}). We then obtain the causal label guidance $ {\tilde o}_{i} $ from all teachers with Equation (\ref{eq:fd_pred}), and the prediction ${\hat o}_{i}$ of the student recommendation model with Equation (\ref{eq:studentpred}). The \textbf{label distillation} loss and the consistency loss are given in Equation (\ref{eq:fd_obj}). Besides two distillation losses, we also employ a widely used recommendation objective $\mathcal{L}_{Rec}$, such as BPR loss \cite{Rendle_Freudenthaler_Gantner_Schmidt} or the sampled softmax loss \cite{Bengio_Senecal_2008}, to train the student recommendation model, \eg,
\begin{align}
\color{myblue} \mathcal{L}_{Rec} =\sum_{(x_i, y_i, o_i) \in \mathcal{D}}-\ln \sigma\left({\hat o}_i - {o}_i\right)+\lambda\|\Theta\|_2^2,
\end{align}
\zsy{where ${\hat o}_i$ denotes the prediction of the student model. ${\hat o}_i$ is basically $P(Y \mid X)$ but approximates $P(Y \mid do(X))$ with causal label distillation (\cf Section \ref{sec:fdaCLD}).} During inference, we use the prediction ${\hat o}_i$ of the student model for recommendation.
Overall, the training objective can be written as:
\begin{align}
	\mathcal{L}_{CausalD} = \mathcal{L}_{Rec} + \lambda_{BDA} \mathcal{L}_{BDA} + \lambda_{FDA} \mathcal{L}_{FDA}. \label{eq:loss}
\end{align}
The training algorithm of the causal multi-teacher distillation is summarized in Algorithm \ref{ag:causalD}. 

\todo{[DONE]+*

6. The formula of L_{rec} is not mentioned in the manuscript. My suggestion is to introduce the loss function, and which distributions that the predictions are made on (P(Y|X) or P(Y|do(X))).

Thank you for your feedback on our paper. According to your suggestions, we have introduced the loss function L_{rec} and add more illustrations of the distributions that our predictions are made on, in Section 3.4.

}

{\SetAlgoNoLine%

	\begin{algorithm}[!t]
		\DontPrintSemicolon
  
  \KwInput{ Recommendation data $\mathcal{D}$ }
  \KwOutput{ Student recommendation model $\{ \phi^*, \psi^* \}$ }
	
Split $\mathcal{D}$ into non-IID subsets $ \{ \mathcal{D}_1, \mathcal{D}_2, \dots, \mathcal{D}_{ K }  \} $

\For{$k = 1$ to $K$}
		{
			Train one teacher on $\mathcal{D}_k$ and obtain encoder $ \phi_k $

		}

Initialize $ \Theta = \{ \phi^*, \psi^*, \mathbf{W}_1, \mathbf{W}_2, \psi_{FDA} \} $
	
	\While{not converged}
	{
	
		Sample a batch $\{(x_i, y_i, o_i)\}_{i=1, \dots, N_b}$ from $\mathcal{D}$ \\
		\For{i = 1 to $N_b$}
		{
			$\mathbf{\hat m}_i = \phi^*( x_i )$
			
			$\mathbf{\tilde m}_{i}^{\circ} = \mathbb{E}_{z_k} \left [ \phi_k(x_i) \right ]$ \Comment*{ Eq.(\ref{eq:bddo})}

			Obtain ${\tilde o}_{i}$ with Eq. (\ref{eq:fd_pred})
			
			Obtain ${\hat o}_i$ with Eq. (\ref{eq:studentpred})

		}
		
		$\mathcal{L}_{BDA} = \sum_i d(\mathbf{\tilde m}_i^{\circ}, \mathbf{\hat m}_i)$ \Comment*{ Eq.(\ref{eq:bddistill})}
		
		$\mathcal{L}_{FDA} = \sum_i \left ( L_{kd}({\tilde o}_i, {\hat o}_i) + L_{ce}({\tilde o}_i, {o}_i) \right )  $  \Comment*{  Eq.(\ref{eq:fd_obj}) }
		
		Obtain $\mathcal{L}_{Rec}$ with BPR \cite{Rendle_Freudenthaler_Gantner_Schmidt}, etc.
		
		Optimize $\Theta$ over Eq. (\ref{eq:loss})

	}

\caption{Causal Multi-teacher Distillation}
\label{ag:causalD}
\end{algorithm}

}%

\subsection{\zsy{Method Analysis}}

\subsubsection{Training Complexity} The computation complexity of CausalD mainly comes from the following processes:
\begin{itemize}[leftmargin=*]
	\item \textit{Teacher Pre-training.} Teachers are trained on non-overlapping subsets of the whole recommendation dataset.	
	Each teacher is with the same neural network architecture as the student. As such, supposing the number of epochs in training the teachers and the student remains the same, the complexity of training all teachers approximates the complexity of training the student.
	\item \textit{Feature encoding.} Supposing that the encoding complexity of the student is $O(E)$, the encoding complexity of the causal distillation framework will be $O(KE + E)$ where $K$ denotes the number of teachers. Note that the teacher encoder and the student encoder have the same structure.
	\item \textit{Prediction.} Supposing that there are $L^*$ layers in $\psi^*$ and the number of hidden units in the $l$-th layer is $d_l^*$, the prediction complexity of the student can be $ O( \sum_{l=1}^{L^*} d_l^* d_{l-1}^* ) $. Supposing that there are $\tilde L$ layers in $\psi_{FDA}$ and the number of hidden units in the $l$-th layer is $ {\tilde d}_l$, the prediction complexity of the teachers can be $   O(\sum_{l=1}^{ {\tilde L}^*} N_b K {\tilde d}_l {\tilde d}_{l-1}) $. 
\end{itemize}
In summary, the computation complexity in causal distillation training can be $ O( KE + E + 2 \sum_{l=1}^{ {\tilde L}^*} N_b K {\tilde d}_l {\tilde d}_{l-1} + \sum_{l=1}^{L^*} d_l^* d_{l-1}^* ) $. In practice, $N_b$ samples can be processed in parallel, and $\sum_{l=1}^{ {\tilde L}^*} N_b K {\tilde d}_l {\tilde d}_{l-1}$ introduces acceptable complexity when ${\tilde d}_l$ and ${\tilde L}$ are small, \ie, $\psi_{FDA}$ is a light-weight network compared to $\psi^*$.
Empirically, with DIN~\cite{Zhou_Zhu_Song_Fan_Zhu_Ma_Yan_Jin_Li_Gai_2018} as the Base model, the multi-teacher pre-training costs around 72s per epoch on the MovieLens dataset. The causal multi-teacher distillation training costs around 157s per epoch. The Base model without distillation costs 64s per epoch under the same experimental setting. 

\subsubsection{Inference Complexity} At inference, solely the student recommendation model is used, and all teachers are discarded. Items are ranked according to the student predictions ${\hat o}$. Therefore, the inference complexity remains the same with the Base model without distillation. Note that the Base model can be any recommendation model with an encoder-predictor architecture. In other words, the causal distillation framework can be plugged into existing recommendation models without affecting the inference efficiency.

\subsubsection{\zsy{Connection between causal label distillation and causal feature distillation}}  \zsy{The underlying connections between causal label distillation and causal feature distillation can be two-fold:}
\begin{itemize}[leftmargin=*]
	\item \zsy{Front-door adjustment and back-door adjustment are two representative causal techniques for deconfounding with unobserved and observed confounders, respectively. We instantiate these two techniques in one unified neural model by identifying mediators and observed confounders. We empirically find that the joint modeling contributes to the overall effectiveness.}
	\item \zsy{As described in Section 3.2, the requirement (ii) of applying front-door adjustment is that no confounders simultaneously affect $X$ and $M$. We further apply the back-door adjustment for deconfounding the estimation of causal effects of $X$ on $M$ (cf. Figure 3 Back-door Adjustment). }
\end{itemize}

\section{Related Works}

%
%
%
%
%

\subsection{Debiased Recommendation} Biases are ubiquitous in the feedback loop (User $\rightarrow$ Data $\rightarrow$ Model $\rightarrow$ User) of recommender systems, including (but are not limited to) exposure bias~\cite{Chen_Feng_Ester_Zhou_Chen_Wang_2018,Liang_Charlin_McInerney_Blei_2016}, popularity bias~\cite{Schnabel_Swaminathan_Singh_Chandak_Joachims_2016}, conformity bias~\cite{Zheng_Gao_Li_He_Li_Jin_2021}, position bias~\cite{Agarwal_Zaitsev_Wang_Li_Najork_Joachims_2019}, and selection bias~\cite{Wang_Zhang_Sun_Qi_2019}. 
One representative line of research to alleviate these biases is to explicitly model or utilize these biases~\cite{DBLP:journals/corr/abs-2109-07946}. Typically, they construct auxiliary models like exposure models \cite{Liang_Charlin_McInerney_Blei_2016,Wang_Liang_Charlin_Blei_2020}, click models \cite{Jin_Fang_Zhang_Ren_Zhou_Xu_Yu_Wang_Zhu_Gai_2020}, and temporal popularity models~\cite{Nagatani_Sato_2017}. Another representative direction is to leverage unbiased data \cite{Liu_Cheng_Dong_He_Pan_Ming_2020,Bonner_Vasile_2018}. For example, \cite{Liu_Cheng_Dong_He_Pan_Ming_2020} collect uniform data by exposing randomly sampled items to each user with a uniform distribution. Undoubtedly, random exposure might hurt users' experiences and the uniform data is expensive to collect. More recently, there is substantial and rapidly-growing research literature studying embracing causal theory in bias alleviation~\cite{DBLP:conf/kdd/WeiFCWYH21,Wang_Feng_He_Wang_Chua_2021,Liu_Cheng_Zhu_Dong_He_Pan_Ming_2021,Wang_Liang_Charlin_Blei_2020,Wang_Feng_He_Zhang_Chua_2021,DBLP:conf/www/HeWC0ZC022,DBLP:conf/kdd/0002G0LLZ0ZYZDT22}.
 Among these, Inverse Propensity Score \cite{Wang_Liang_Charlin_Blei_2018,Yang_Cui_Xuan_Wang_Belongie_Estrin_2018} is one of the most general and representative methods in reducing confounding effects. In essence, IPS re-weights training samples based on the estimated bias-aware propensity scores. However, most existing deconfounding works (including IPS) assume unconfoundedness \cite{Rubin_1978}, which means the confounders are observable to the model. However, in real-world recommender systems, confounders are too various to be modeled in one model, and a considerable portion of them are even unobserved by the system. 
Different from these works, we leverage front-door adjustment technique from causality to handle unobserved confounders, and derive an efficient approximation of it using causal multi-teacher distillation. 

It is noteworthy that we differ from fairness-related works \cite{Li_Chen_Fu_Ge_Zhang_2021} both in motivation and technique. Our major focus is the model training bias that enlarges the performance heterogeneity of different user groups. As shown in Figure \ref{fig:longtail_base}, when users in different groups are trained all together, and trained independently, a small portion of users benefit more during the co-training of all users. Note that performance heterogeneity still exists without co-training, and such heterogeneity is \textit{\textbf{not}} our major focus. Fairness-related works typically penalize the prediction expectation (such as the positive prediction rate) of different user groups, ensuring the universal equality. For example, \cite{Li_Chen_Fu_Ge_Zhang_2021} work in a post-processing manner with recommended Top-K items and their ranking scores as input. They try to ensure equal performance of different user groups by solving a 0–1 integer programming problem. Different from existing works, we use causal techniques to deconfound model training.

\subsection{Knowledge Distillation}

\vpara{Multi-teacher Distillation.}
Knowledge distillation is a powerful tool for transferring knowledge between networks. The transferred knowledge takes the form of soft labels~\cite{DBLP:journals/corr/HintonVD15}, intermediate features~\cite{Zagoruyko_Komodakis_2017,Heo_Kim_Yun_Park_Kwak_Choi_2019}, and relations~\cite{Park_Kim_Lu_Cho_2019,Peng_Jin_Li_Zhou_Wu_Liu_Zhang_Liu_2019}. More recently, there is a rapidly-growing research literature studying on multi-teacher distillation \cite{Shen_He_Xue_2019,Tan_Ren_He_Qin_Zhao_Liu_2019}. Among them, MEAL \cite{Shen_He_Xue_2019} is one of the early works that train multiple teachers. It measures the consistency between the selected teacher and the student using adversarial learning. \cite{Tan_Ren_He_Qin_Zhao_Liu_2019} encapsulates multi-teacher distillation into multi-task learning, and trains one teacher per task. EnsCTR \cite{Zhu_Liu_Li_Lai_He_Chen_Zheng_2020} makes the first attempt to ensemble multiple teachers for click-through-rate prediction. They devise the Teacher Gating network to assign weights to different teachers based on their predictions. To the best of our knowledge, our work is the first to consolidate causality into multi-teacher distillation.

\vpara{Knowledge Distillation in Recommendation.} In recommender systems, there are some works that exploit a privileged teacher to enhance the student that is used for serving/inference \cite{Wang_Zhang_Wu_Ma_Hong_Wang_2021,Lee_Choi_Lee_Shim_2019}. The privilege of teachers mainly falls into two groups: 1) high-capacity in model's architecture \cite{Kang_Hwang_Kweon_Yu_2021,Wang_Yin_Chen_Huang_Wang_Zhao_Hung_2020} where related works typically train a lighter-weight student for efficiency; and 2) external knowledge such as user reviews \cite{Chen_Zhang_Xu_Qin_Zha_2019}, on-device features \cite{Yao_Wang_Jia_Han_Zhou_Yang_2021}, and unbiased uniform data \cite{Liu_Cheng_Dong_He_Pan_Ming_2020}. Largely different from these works, we construct unprivileged teachers, which have the same architecture as the student, and each of them is solely on a subset of the training data used to train the student. Though these teachers are unprivileged, we distill causal effects from them and accordingly train a debiased student without external unbiased data, which constitutes a novel paradigm for leveraging distillation in recommender systems.

\begin{table}[!t]
\caption{Statistics of the datasets.}
\centering
\small
\begin{tabular}{l cc cc}
\toprule
Dataset & \#Users & \#Items & \#Interactions & \#Density  \\
    \midrule
    MovieLens         & 6, 040 & 3, 900 & 1, 000, 209 & 0.04246 \\ 
    Amazon          & 459, 133 & 313, 966 & 8, 898, 041 & 0.00063 \\ 
    Alipay   & 400, 594 & 19, 503 & 38, 710, 494 & 0.00495 \\
    \bottomrule
\end{tabular}%
    \label{tab:staData}
\end{table}
%

\begin{table*}[h]
\centering
    \caption{Overall performance and performance heterogeneity comparison with DIN as the base model. 
$p$-value is from the two-sided test between CausalD and the best-performing baseline.
    Higher values are better except for Heterogeneity.}
    
\small
{\setlength{\tabcolsep}{0.48em}\renewcommand{\arraystretch}{1.}\begin{tabular}{ll cccccccccc}

\toprule

  Datasets  & Metric & \textbf{DIN} & KD  & IPS &	CausalRec &  DeRec & DebiasD & EnsCTR & MEAL &   CausalD & $p$-value  \\
    \midrule
   
  \multirow{6}{*}{MovieLens} 
   &  AUC  & 	0.8185   &  0.8187	   &   0.8034	&	0.8176 &  0.8227	&	0.8227   &  \underline{0.8257}	&  0.8233		 &  \textbf{0.8329}	  &	  1.36e-07 \\
   &  R@5  & 	0.3102   &  0.3104	   &   0.3066	&	0.3112 &  0.3260	&	0.3231   &  \underline{0.3354}	&  0.3213		 & \textbf{0.3556}	  &	   3.55e-06\\
   &  R@10 & 	0.4682   &  0.4763	   &   0.4652	&	0.4751 &  0.4892	&	0.4831   &  \underline{0.4930}	&  0.4837		 &  	\textbf{0.5170}    &	   4.67e-05 \\
   &  NDCG@5 & 	0.2378   &  0.2382	   &   0.2362	&	0.2400 &  0.2507	&	0.2485  &  \underline{0.2589}	&  0.2492		 &  	\textbf{0.2755}    &	    1.15e-06 \\
   &  NDCG@10 & 0.3145	&  0.3187	   &   0.3136	&	0.3192 &  0.3300	&	0.3258  &  \underline{0.3355}	&  0.3281		 & \textbf{0.3538}	  &	   5.53e-06 \\
   & Heterogeneity $\downarrow$   & 0.7291  & 0.4533  &   \underline{0.3056} & 0.7529& 0.7893  & 0.6889  & 0.9531  & 0.7773  & \textbf{0.1086}  & 1.54e-03 \\ 
    \midrule

    \multirow{6}{*}{Amazon} 
   &  AUC  & 	0.8873   &   0.8704	   &   	0.8788	&	0.8899 & 0.8908 	& 	0.8954	&   \underline{0.8983}	&   0.8867	   	 & \textbf{0.9027}	  &	   2.30e-07 \\
   &  R@5  & 	0.5644   & 0.5336 	   &   	0.5555 	&	0.5790 &	0.5795 &  	0.5886	&   \underline{0.5911}	&   0.5698	   	 & \textbf{0.6069}	  &	   3.24e-04 \\
   &  R@10 & 	0.6854   & 0.6515 	   &   	0.6760	&	0.6956 & 0.6956 & 	0.7060	&   \underline{0.7087}	&   0.6863	   	 & \textbf{0.7216}	  &	  1.52e-05  \\
   &  NDCG@5 & 	0.4794   & 0.4516 	   &   	0.4706	&	0.4934 & 	0.4939 & 	0.5007	&   \underline{0.5038}	&   0.4857	   	 &  	\textbf{0.5190} &	   3.71e-05 \\
   &  NDCG@10 & 0.5384	 & 0.5092 	   &   	0.5294	 &	0.5503 &	0.5506 &  	0.5580	&   \underline{0.5612}	&   0.5426      	 &  	\textbf{0.5751} &	  2.58e-05  \\
   & Heterogeneity $\downarrow$   & 2.4567& 1.6847 & \underline{1.4277} & 1.8765 &  1.6241 &  1.4747  & 1.5169  & 2.2083 & \textbf{1.3803}  & 1.89e-02 \\

    \midrule
    \multirow{6}{*}{Alipay}  
   &  AUC    & 	0.7691   & 0.7615	   &   0.7623	 &	0.7603 &		0.7738 &     0.7712	&   0.7700	&   \underline{0.7749}	   	 & \textbf{0.7777} 	  &	3.08e-05    \\
   &  R@5    & 	0.1669   & 0.1727	   &   0.1778	 &	0.1792 & 	0.2039 &    0.2057	&   0.1831	&   \underline{0.2343}	   	 &\textbf{0.2547} 	  &	5.73e-08    \\
   &  R@10   & 	0.3518   & 0.3682	   &   \underline{0.3953}  &	0.3772 &		0.3947 &    0.3675	&   0.3689	&   0.3938	   	 &\textbf{0.4457} 	  &	2.22e-06    \\
   &  NDCG@5 & 	0.1186   & 0.1221	   &   0.1237	 &	0.1255 &		0.1474 &     0.1532	&   0.1307	&   \underline{0.1745}	   	 &\textbf{0.1851} 	  &	1.53e-06    \\
   &  NDCG@10 & 0.2076   & 0.2165	   &   0.2289	 &	0.2214 &		0.2407 &     0.2314	&   0.2204	&   \underline{0.2517}	   	 &\textbf{0.2779} 	  &	1.86e-05    \\
   & Heterogeneity $\downarrow$  & 2.3419  & 2.3383 &1.8455  & \underline{1.5405} & 2.4671 & 2.1373  & 2.4441  & 2.1125  & \textbf{1.0207}  & 3.39e-03 \\ 
    \bottomrule
\end{tabular}}
    \label{tab:comparison_DIN}
\end{table*}
%

\section{Experiments}

%
%

%
%
%

\todo{[DONE][Exp]+++***

2. The authors argued that a wise goal is to eliminate the model bias while maintaining the natural heterogeneity. The experimental results cannot prove that the goal is achieved. In the manuscript, the authors adopted Heterogeneity metric. However, this metric only evaluates the mixture of model biases and the natural heterogeneity. Is there any specific metric for model bias rather than evaluating mixture of model biases and the natural heterogeneity?

We appreciate your valuable suggestion and would like to express our gratitude for your input. As a result of your feedback, we have made revisions to our research paper. Specifically, in Section \ref{sec:problemformu}, we have formally defined the concept of bias-amplified performance heterogeneity. To evaluate this concept, we assess the performance heterogeneity of various user groups under unified training and independent group-wise training. Under the latter approach, users from a specific group do not experience bias from other groups, and the performance heterogeneity could approximate the natural heterogeneity. We calculate the difference between the two heterogeneities to obtain the bias-amplified performance heterogeneity. This revised approach aligns with the concept introduced in the Introduction and provides a more consistent evaluation. To reflect these changes, we have updated the performance heterogeneity scores in Table 3 and Table 4, and provided a revised illustration of the Heterogeneity metric in Section 5.1.

}

We conduct experiments on three real-world datasets to answer the following research questions:

\begin{itemize}[leftmargin=*]
	\item \textbf{RQ1:} How does \textit{CausalD} perform as compared to baselines concerning multi-teacher distillation, debiased distillation, and debiased recommendation? 
	\item \textbf{RQ2:} Does \textit{CausalD} mitigate the performance heterogeneity problem amplified by confounders that are observed or unobserved by the model?
	\item \textbf{RQ3:} How do different components (the back-door and front-door adjustment loss) and different hyper-parameter settings (\eg, loss coefficients, number of teachers) affect the performance of CausalD?
\end{itemize}

\subsection{Experimental Setup}

\vpara{Dataset Description.} We use two public benchmark datasets, \ie, MovieLens and Amazon Product Review, and a challenging large-scale industrial dataset, \ie, AliPay. Dataset statistics are summarized in Table \ref{tab:staData}.
\begin{itemize}[leftmargin=*]
	\item \textbf{Amazon.} We take the Book category from the Amazon product review dataset\footnote{http://jmcauley.ucsd.edu/data/amazon/} for evaluation. We follow \cite{Zhou_Zhu_Song_Fan_Zhu_Ma_Yan_Jin_Li_Gai_2018,Wang_He_Wang_Feng_Chua_2019} to regard the interacted items for a given user as positive items and pair the user with randomly sampled negatives that the user did not consume before for training models.
	\item \textbf{MovieLens.} We use the widely used MovieLens-1M\footnote{https://grouplens.org/datasets/movielens/1m/} version for evaluation. We generate positive and negative pairs in a similar manner as  the Amazon dataset.
	\item \textbf{AliPay.} AliPay dataset is collected from one of the world's largest mobile payment platforms, \ie, AliPay. Wherein, the applets like mobile recharge services and government services are viewed as items. In AliPay, we regard the items that have been clicked as positives. Items that are observed but not clicked are treated as negatives.
\end{itemize}
To ensure data quality, we take the 16-core setting where each user/item has at least sixteen interactions.

\vpara{Baselines.}  To demonstrate the effectiveness, we compare CausalD with the following baselines:

\begin{itemize}[leftmargin=*]
	\item \textbf{KD \cite{DBLP:journals/corr/HintonVD15}: } This model denotes the oracle knowledge distillation. We train a single teacher on all users.
	\item \textbf{IPS \cite{Yang_Cui_Xuan_Wang_Belongie_Estrin_2018}:}. It is a representative causal recommendation method. In this work, we firstly measure $P(ug|i)$ which denotes the conditional probability of heterogeneous user groups $ug$ given items $i$. Then, for a given data sample $(u,i)$, we use the corresponding $p(ug|i)$ as the propensity score. We follow \cite{Yang_Cui_Xuan_Wang_Belongie_Estrin_2018} to reduce propensity variance using the propensity clipping technique. The clipping threshold is searched in $\{1/2, 1/3, ..., 1/10\}$.
	\item \textbf{CausalRec~\cite{liang2016causal}:} This method employs causal inference to correct the exposure bias and technically constructs an exposure model such that each click is weighted by the inverse exposure probability predicted by the model.
	\item \textbf{Deconfounded Rec (DeRec)~\cite{Wang_Liang_Charlin_Blei_2018}:} The deconfounded recommendation technique uses Poisson factorization to infer confounders, and augments the base recommendation model to correct the confounding.
	\item \textbf{DebiasD \cite{Lukasik_Bhojanapalli_Menon_Kumar_2021}:} It is one state-of-the-art method that debiases single-teacher distillation at the class level. We extend the framework to the multi-teacher distillation by assigning adaptive mixing weights to multiple teachers. 
	\item \textbf{EnsCTR \cite{Zhu_Liu_Li_Lai_He_Chen_Zheng_2020}:} It is one state-of-the-art multi-teacher distillation framework in the literature of recommendation.
%
	\item \textbf{MEAL \cite{Shen_He_Xue_2019}:} This baseline is a state-of-the-art ensemble distillation in the generic domain. It incorporates discriminators to evaluate to what extent the student's output differs from multiple teachers. 
\end{itemize}

%

\begin{table*}[h]
\centering
    \caption{Recommendation performance and performance heterogeneity comparison with DeepFM as the base model. 
    }
\small
{\setlength{\tabcolsep}{0.4em}\renewcommand{\arraystretch}{1}\begin{tabular}{ll cccccccccc}
\toprule

  Datasets  & Metric & \textbf{DeepFM} & KD  & IPS & 	CausalRec & DeRec & DebiasD & EnsCTR & MEAL & CausalD & $p$-value   \\
    \midrule
   
    \multirow{6}{*}{MovieLens} 
   &  AUC  & 	0.8142   &   0.8234	   &   	0.8113	&	0.8172 & 	0.8142 & 	0.8231	&   \underline{0.8241}	&   0.8210	  	 & \textbf{0.8320}	  &	  4.68e-03  \\
   &  R@5  & 	0.2867   & 0.3047 	   &   	0.2867	&	0.3021 &		0.2902 &		0.3064	&   \underline{0.3142}	&   0.3039	  	 & \textbf{0.3249}	  &	  3.88e-02  \\
   &  R@10 & 	0.4463   & 0.4633 	   &   	0.4469	&	0.4537 & 	0.4509 &		0.4685	&   \underline{0.4787}	&   0.4691	   	 & \textbf{0.4916}	  &	 2.66e-02   \\
   &  NDCG@5 & 	0.2230   & 0.2362 	   &   	0.2188	&	0.2338 & 	0.2241 &		0.2348	&   \underline{0.2420}	&   0.2354	   	 &  	\textbf{0.2481} &	 1.47e-02   \\
   &  NDCG@10 & 0.3005	 & 0.3129 	   &   	0.2967	&	0.3072 &	0.3019 & 	0.3134	&   \underline{0.3216}	&   0.3152      	 &  	\textbf{0.3290} &	 2.62e-02   \\
   & Heterogeneity $\downarrow$   & 7.1753 &5.8783  & \textbf{5.4983} & 5.7853 &5.7893   & 6.4233  &6.1513 &5.9233 & \underline{5.6333}  & - \\ 
    \midrule

    \multirow{6}{*}{Amazon} 
   &  AUC  & 	0.8941   &   0.8827	   &   	0.8886	&	0.8957 & 0.8956	& 	0.8953	&   0.8913	&   \underline{0.8963}	   	 & 		\textbf{0.8977}	  & 3.39e-03    \\
   &  R@5  & 	0.5693   & 0.5723 	   &   	0.5568	&	0.5730 &	0.5718 &	0.5743	&   \underline{0.5855}	&   0.5760	   	 & 		\textbf{0.6039}	  &	6.62e-06	    \\
   &  R@10 & 	0.6877   & 0.6880 	   &   	0.6775	&	0.6918 &	0.6904 &	0.6923	&   0.6934	&   \underline{0.6947}	   	 & 		\textbf{0.7087}	  &	1.89e-03	    \\
   &  NDCG@5 & 	0.4838   & 0.4876 	   &   	0.4721	&	0.4869 &	0.4865 & 	0.4886	&   \underline{0.5018}	&   0.4904	   	 &  		\textbf{0.5191} &	1.49e-04	    \\
   &  NDCG@10 & 0.5416	 & 0.5440 	   &   	0.5310	&	0.5449 & 0.5444 &	0.5461	&   \underline{0.5544}	&   0.5483     	 &  		\textbf{0.5716} 	&	6.94e-03    \\
   & Heterogeneity $\downarrow$   & 0.2224 & \underline{0.1422}  & 0.1082  & 0.1840 & 0.2350 & 0.1988  &0.2376  & 0.1414  & \textbf{0.0762}  & 5.10e-04 \\ 
    \midrule
    \multirow{6}{*}{Alipay} 
   &  AUC  & 	0.8018   &   0.8006	   &   	\underline{0.8058}&	\underline{0.8058} & 	0.8016 & 	0.8017	&   \textbf{0.8077}	&   0.8016	   	 & 	0.8031  &	  -  \\
   &  R@5  & 	0.2094   & 0.2227 	   &   	0.2253	&	0.2326 & 	0.2281 & 	0.2274	&   \underline{0.2612}	&   0.2292	   	 & \textbf{0.2909}	  &	  2.61e-04  \\
   &  R@10 & 	0.4177   & 0.4263 	   &   	0.4276	&	0.4314 &	0.4298 & 	0.4313	&   \underline{0.4492}	&   0.4257	   	 & \textbf{0.4657}	  &	  2.16e-02 \\
   &  NDCG@5 & 	0.1491   & 0.1615 	   &   	0.1626	&	0.1674 &	0.1633 & 	0.1637	&   \underline{0.1939}	&   0.1661	   	 &  	\textbf{0.2207} &	  1.63e-05  \\
   &  NDCG@10 & 0.2501	 & 0.2603 	   &   	0.2607	&	0.2639 &	0.2614 & 	0.2626	&   \underline{0.2848}	&   0.2614     	 &  	\textbf{0.3053} &	  1.06e-02  \\
   & Heterogeneity $\downarrow$   &2.0041 &2.0877  & 1.6609  & 1.6577 & 1.8715 & 1.6355 & 2.3579  & \underline{1.5945}  & \textbf{1.5747}  & 2.31e-02\\ 
    \bottomrule
\end{tabular}}
    \label{tab:comparison_DeepFM}
\end{table*}
%

\vpara{Implementation Details and Evaluation.} Since \textit{CausalD} is designed as model-agnostic, we instantiate CausalD and baselines on two representative base recommenders, \ie, Deep Interest Network (DIN)~\cite{Zhou_Zhu_Song_Fan_Zhu_Ma_Yan_Jin_Li_Gai_2018} and DeepFM~\cite{Guo_Tang_Ye_Li_He_2017}. All models except for KD, which is a single-teacher distillation baseline, share the same pre-trained multiple teachers. We optimize all models with Adagrad \cite{Duchi_Hazan_Singer_2011} as the optimizer and with the default choice of learning rate (\ie, $0.01$). All models are with batch size 4096, and user/item embedding size 8. For DIN and models constructed upon it, we consider the last 50 items interacted to form the user sequence. For loss coefficients of all models that introduce additional loss functions (including $\mathcal{L}_{BDA}$ and $\mathcal{L}_{FDA}$ in CausalD), we search in the range of $\{ 0.01, 0.1, 1, 10 \}$.  We run CausalD and the best-performing baseline for five times independently, and conduct two-sided tests.

 It is known that industrial recommender systems typically consist of two phases, \ie, the matching phase (also known as deep candidate generation) and the ranking phase. In the literature, evaluation strategies for these two lines of works are different. 
For example, ranking models solely sort a small group of candidates rather than the whole item gallery. Therefore, ranking models can incorporate complex feature interactions between items and users. 
In this paper, we focus on the ranking phase and obey the next-item recommendation protocol to evaluate the performance of models, which has been widely used in~\cite{Zhou_Zhu_Song_Fan_Zhu_Ma_Yan_Jin_Li_Gai_2018,He_Liao_Zhang_Nie_Hu_Chua_2017}. Specifically, we chronologically order the interacted items for each user to have a behavior sequence, and leave the last element of the behavior sequence for testing. The testing target item is paired with 100 randomly sampled negative items that the user has not consumed. 
To have a comprehensive analysis of models, we employ various \textit{evaluation metrics}, including Area Under the ROC Curve (AUC), Normalized Discounted Cumulative Gain (NDCG), and Recall. We report metrics computed on the top 5/10 items. Besides these widely used metrics, we also evaluate the performance heterogeneity of different users. \zsy{The \textit{\textbf{Heterogeneity}} metric is the mean of five bias-amplified performance heterogeneity scores computed according to Equation \ref{eq:training_hetero} over five metrics (AUC, R@5, R@10, NDCG@5, NDCG10).}
The lower values, the better.

\todo{[DONE]++*

6. The performance improvement of CausalD on DeepFM seems to be less consistent than CausalD on DIN. I suggest the authors provide more discussions on this phenomenon.

Thanks for your suggestion. We have added more discussions on this phenomenon in Section 5.2 as follows:

We also notice that with DeepFM as the base model, CausalD cannot beat other baselines in a few cases while CausalD with DIN as the base model could consistently achieve the best results. We attribute this phenomenon to the two parallel components, \ie, the FM component and the DNN component, of DeepFM. On the one hand, the causal feature distillation that operates on intermediate representations is mostly affecting the DNN component while the FM component might still suffer from training biases. On the other hand, two parallel components would result in two independent predictions while their relative importance might be hard to determine under causal label distillation. Still, we observe that CausalD over DeepFM could achieve the best results in most cases, demonstrating the model-agnostic effectiveness of CausalD.

}

\todo{[DONE][Exp]++*

2. In Deconfounding Analysis (RQ2) and Figure 4, the authors discuss the performance of one base model, and two proposed architectures. Readers would be interested in how causal recommendation baseline DeRec and debiased distillation baseline DebiaseD would perform. I hereby suggest the authors add such results and more discussions.

We appreciate your valuable suggestion, which has helped us improve our paper significantly. In response to your feedback, we made two key changes to the paper. Firstly, we have included the results of causal recommendation baseline DeRec and debiased distillation baseline DebiaseD in Figure 4. Secondly, we have expanded the discussions in Section 5.3.

}

\begin{figure*}[!t] \begin{center}
\begin{subfigure}{.48\textwidth}
    \includegraphics[width=\columnwidth]{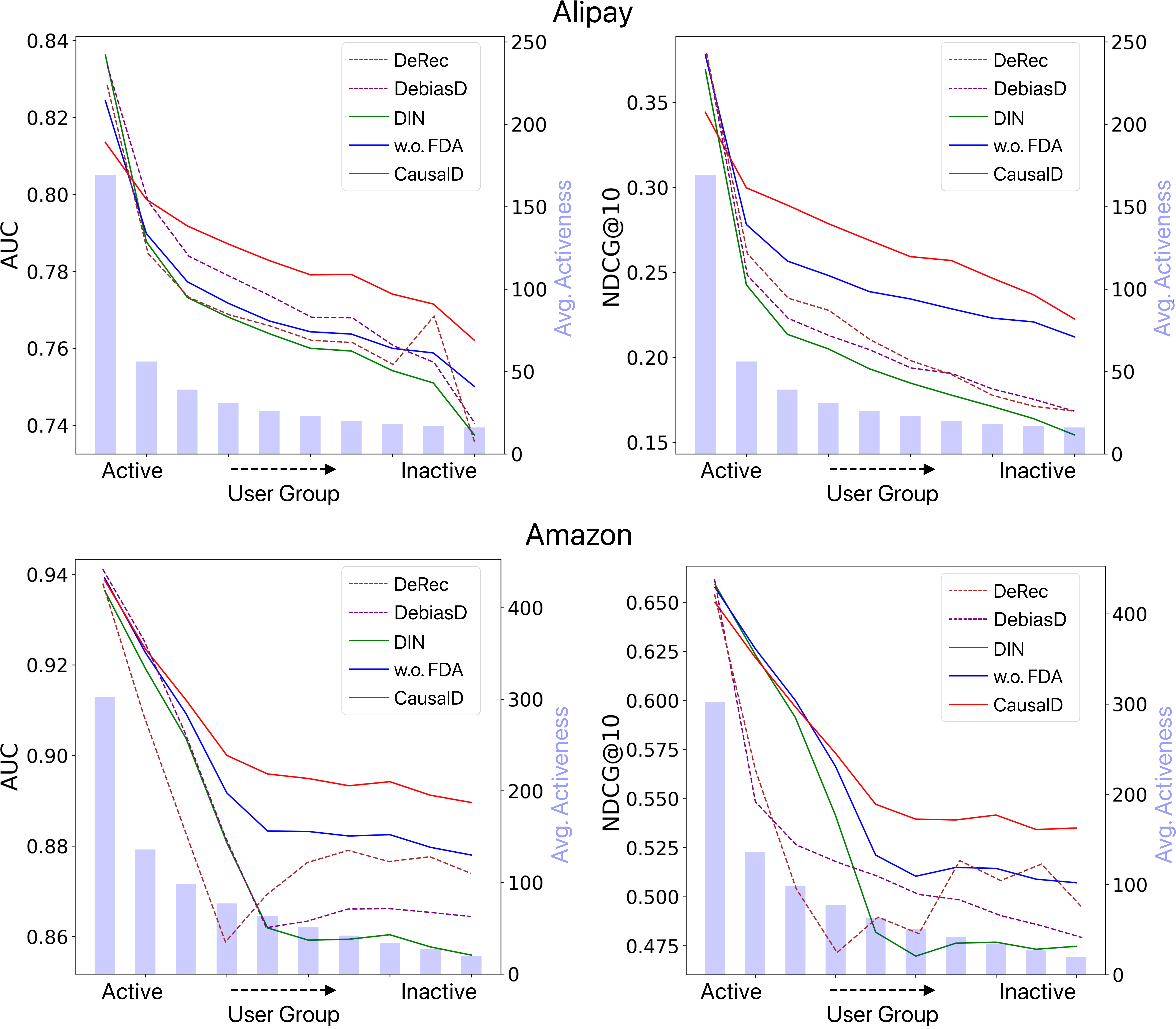}
    \caption{
    	Observed confounder: \textit{user activeness}.
	}
\label{fig:debias_active}
\end{subfigure}
\begin{subfigure}{.48\textwidth}
    \includegraphics[width=\columnwidth]{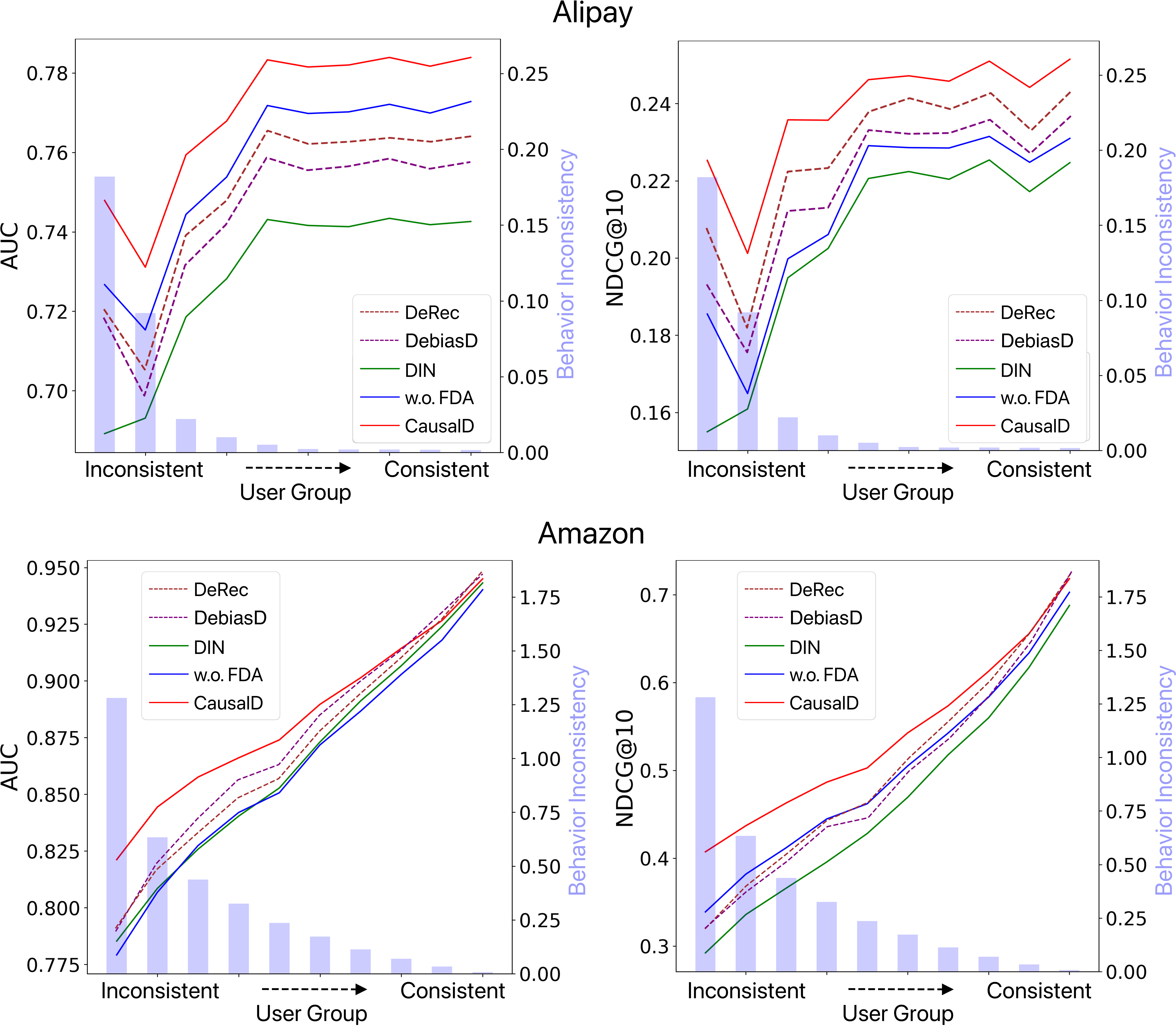}
    \caption{
        Unobserved confounder: user \textit{behavior consistency}.
	}
\label{fig:debias_pop}
\end{subfigure}
    \caption{
    	Performance heterogeneity across different user groups \wrt two confounders, \ie, user activeness, and behavior consistency. The performance heterogeneity over user groups is largely reduced compared to the base model DIN~\cite{Zhou_Zhu_Song_Fan_Zhu_Ma_Yan_Jin_Li_Gai_2018}. 
    	}
    \label{fig:citeseer}
\end{center} \end{figure*}

\subsection{Overall Performance (RQ1)} \label{sec:overallPerform}

We run CausalD and all baselines on three datasets. For each model, we run it five times and take the average averaged results. The results with DIN and DeepFM as base models are shown in Table \ref{tab:comparison_DIN} and Table \ref{tab:comparison_DeepFM}, respectively. Higher values are better for AUC, NDCG, and Recall except for Heterogeneity. From the comprehensive results, we have the following observations:

\begin{itemize}[leftmargin=*]
	\item In summary, CausalD achieves the best performance and the least performance heterogeneity in most cases. For example, CausalD outperforms DIN and the best performing baseline EnsCTR by 15.85\% and 6.41\% (NDCG@5), and reduces the performance heterogeneity by 8.14\% and 11.09\% on the MovieLens dataset. The performance gains are statistically significant with $p$-value $< 0.05$ under two-sided tests\footnote{\url{https://docs.scipy.org/doc/scipy/reference/generated/scipy.stats.ttest_ind.html}}. Improving the overall performance while reducing the heterogeneity probably indicates that tail users benefit more with CausalD compared to other baselines. We further reveal the rationality of this analysis in Section \ref{sec:cra}. On the contrary, some other baselines (such as IPS) achieve low heterogeneity and low performance at the same time, which means that they might hurt the performance of head users. The improvements are consistent across different recommendation architectures and different datasets, showing the merits of CausalD being model-agnostic and domain-agnostic.
	 On the larger-scale dataset, Alipay, the improvements are even more substantial. In essence, these results demonstrate the necessity of reducing confounding effects for recommender systems, and the effectiveness of the causal distillation framework.
	\item The oracle distillation framework (KD) obtains almost no gains \wrt the base models. This indicates that when the model capacity and the training data remain the same, the unprivileged teacher might not be very useful in recommender systems. This method can reduce the heterogeneity to a limited extent. We attribute this result to that soft labels produced by the teacher can probably convey some confidence judgments on the predictions for different users. For example, active users might get smoother soft labels than inactive ones. 
	\item IPS reduces the performance heterogeneity in many cases by considering user-specific propensity scores. However, it obtains almost no gains on recommendation performance and sometimes does harm on MovieLens and Amazon with DIN as the base model. One reasonable explanation is that IPS largely sacrifices the performance of head users for the performance of long-tail users. Compared to IPS, exposure-aware causal baselines CausalRec and DeRec can consistently improve the Base model. By correcting the exposure bias, these models are potentially less affected by the spurious correlations between users and frequently exposed items. Tail users could benefit from the correction since they suffer more from spurious correlations as illustrated in the Introduction. However, they might not effectively handle other ubiquitous confounders (\eg, user activeness, item popularity) beyond exposure.
		The debiased distillation baseline DebiasD can also obtain some performance gains and reduces the performance heterogeneity to some extent. It learns whether we should trust teachers and how differently we should trust different teachers in distillation. But still, its results are inferior to CausalD in most cases.
	\item Muti-teacher distillation baseline EnsCTR achieves good performance compared to other baselines. It leverages multi-teachers to maximize the recommendation performance. However, the heterogeneity is mostly worse than other baselines and even worse than the base model in some cases, which probably means EnsCTR further sacrifices the performance of tail users. The performance of MEAL seems to be less consistent in different experimental settings. Unstable training caused by adversarial learning is also found in many other research fields \cite{Salimans_Goodfellow_Zaremba_Cheung_Radford_Chen_2016}.
	\item \zsy{We also notice that with DeepFM as the base model, CausalD cannot beat other baselines in a few cases while CausalD with DIN as the base model could consistently achieve the best results. We attribute this phenomenon to the two parallel components, \ie, the FM component and the DNN component, of DeepFM. On the one hand, the causal feature distillation that operates on intermediate representations is mostly affecting the DNN component while the FM component might still suffer from training biases. On the other hand, two parallel components would result in two independent predictions while their relative importance might be hard to determine under causal label distillation. Still, we observe that CausalD over DeepFM could achieve the best results in most cases, demonstrating the model-agnostic effectiveness of CausalD. }
\end{itemize}

\begin{figure*}[!t] \begin{center}
    \includegraphics[width=\textwidth]{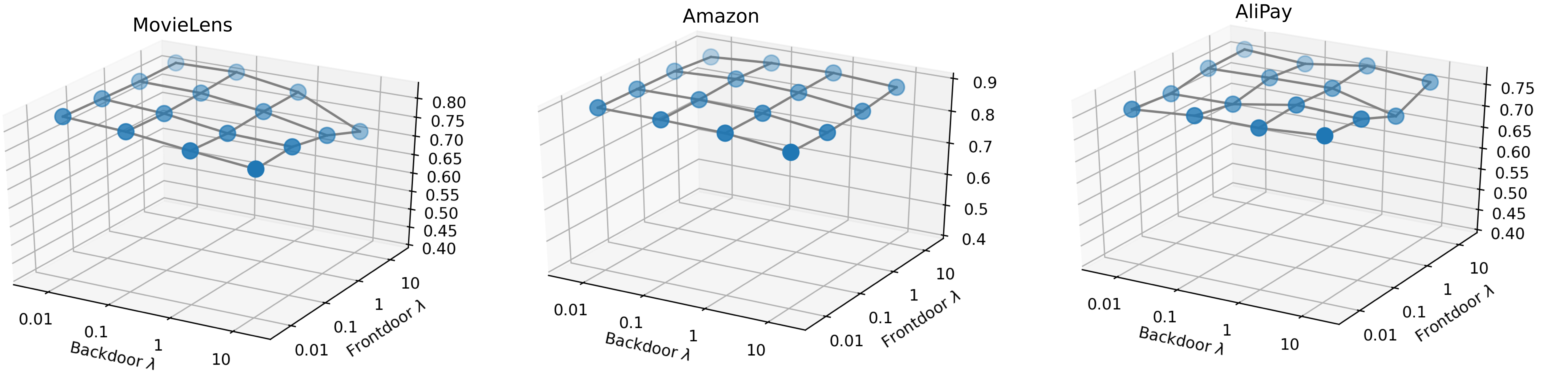}
    \caption{
    	Recommendation performance with loss coefficients ($\lambda_{BDA}$ and $\lambda_{FDA}$) varying in range $\{ 0.01, 0.1, 1, 10 \}$. 
	}
\label{fig:lambda}
\end{center} \end{figure*}

\subsection{Deconfounding Analysis (RQ2)}  \label{sec:cra}

We are interested in whether CausalD reduces the performance heterogeneity of users by deconfounding and improving the performance of tail users with minor performance sacrifice of head users. In this regard, we report the recommendation performance of models on different user groups split \wrt two confounders, \ie, 1) \textit{user activeness}, which is observed by the model, and 2) users' \textit{behavior consistency} on popular items, which is unobserved by the model. We evaluate CausalD, a variant of CausalD without loss $\mathcal{L}_{FDA}$ (w.o. FDA), and the Base DIN model.
Figure \ref{fig:debias_active} and Figure \ref{fig:debias_pop} show the results. We plot both the recommendation performance (the lines) and the values of confounders (the bars), including user activeness and behavior inconsistency. According to the results, we have the following observations: 
\begin{itemize}[leftmargin=*]
	\item Overall, these results further show the effectiveness of CausalD on improving recommendation performance and reducing performance heterogeneity of different users.
	\item By analyzing the performance over different groups in detail, we observe that CausalD improves the recommendation performance by mainly and largely improving the results of tail users that suffer from low-quality recommendation while mostly preserving the recommendation quality of head users. These results basically demonstrate that CausalD reaches the major goal of this work, \ie, to alleviate model training bias that is in favor of head users, and to reduce performance heterogeneity by mainly improving the performance of tail users that originally benefit little from the co-training of all users. This merit can be potentially of great value in real-world recommender systems where tail users are the majority and cold-start problems are ubiquitous.
	\item \zsy{Although the DeRec causal recommendation baseline and the DebiasD debiased distillation baseline can provide some relief for performance heterogeneity issues in certain scenarios, there remains a significant disparity in bias alleviation performance compared to causalD. These results are consistent with those outlined in Section \ref{sec:overallPerform} and the analytical observations was comparable.}
	\item CausalD can successfully handle confounders (\eg, behavior consistency) that are unobserved to the model. This is technically critical merit of the CausalD framework. In real-world scenarios, we cannot observe all confounders that cause the spurious correlations of historical behaviors and the next behaviors, resulting in inevitable performance heterogeneity and non-robustness. Dealing with unobserved confounders is enticing but challenging, and this work demonstrates the front-door adjustment might be a promising technique for recommender systems.
	\item \zsy{We also notice that, in some cases, the performance of CausalD on head users becomes worse than the Base model. This phenomenon can be intuitive since by alleviating training bias brought by potential confounders, the model might capture less biased prediction shortcuts that are in favor of the majority user group. For example, some active users might click a lot of popular items disregarding their inherent interests, leading to biased records. Debiased models might recommend items that are less affected by the item popularities, and thus probably hurting the performance of the above users. Still, CausalD achieves consistently better performance for all user groups in most cases. We leave user-wise bias exploitation as a potential future work. }
\end{itemize}

esults of causal recommendation baseline DeRec and debiased distillation baseline DebiaseD in Figure 4


\todo{[DONE] +*

5. In Deconfounding Analysis (RQ2) and Figure 4, the authors reveal that the proposed method significantly improves the performance of tail users, and alleviates performance heterogeneity, which is promising. I also notice that, in some cases, the performance of head users becomes worse. I suggest that the authors provide more insights and illustrations on such a phenomenon.

Thanks for your suggestion. We have added more discussions of this phenomonen in Section 5.3 as follows:
We also notice that, in some cases, the performance of CausalD on head users becomes worse than the Base model. This phenomenon can be intuitive since by alleviating training bias brought by potential confounders, the model might capture less biased prediction shortcuts that are in favor of the majority user group. For example, some active users might click a lot of popular items disregarding their inherent interests, leading to biased records. Debiased models might recommend items that are less affected by the item popularities, and thus probably hurting the performance of the above users. Still, CausalD achieves consistently better performance for all user groups in most cases. We leave user-wise bias exploitation as a potential future work. 
}

\todo{[DONE] [Exp]++*

1. The ablation study reveals the necessity of heterogenous teachers, front-door adjustment, and back-door adjustment with DIN as the base model. I am interested in whether similar observations exist in DeepFM, and suggest the authors add more ablation studies.

Thanks for your valuable suggestions. We have added the ablation study with DeepFM as the base model. According to the results listed in Table \ref{tab:ablation_dfm}, we have similar observations that are consistent with those presented in Table \ref{tab:ablation}.

}

\begin{table}[!t]
\centering
    \caption{Ablation study with DIN as the base model \wrt three critical components: 1) front-door adjustment (w.o. FDA $\rightarrow$ CausalD); 2) back-door adjustment (Base $\rightarrow$ w.o. FDA); and 3) heterogeneous teachers (Homogeneous (Ho) Teachers $\rightarrow$ CausalD).}
    \small
{\setlength{\tabcolsep}{0.4em}\begin{tabular}{l cccccc}
\toprule
 &\multicolumn{3}{c}{ Amazon } &\multicolumn{3}{c}{ Alipay } \\
\midrule
Model
   & AUC & NDCG & Recall & AUC & NDCG & Recall \\
    \midrule
	 DIN (Base) &	 0.8873 &	 0.5384 &	 0.6877 & 	 0.7691 & 	 0.2076 & 	 0.3518 \\
	 \midrule
	 Ho Teachers &	 0.8869 &	 0.5429 &	 0.6883 & 	 0.7723 & 	 0.2289 & 	 0.3953 \\
		w.o. FDA &	0.8952 &	 0.5527 &	0.7008 & 	 0.7727 & 	 0.2238 & 	0.4168 \\
	 CausalD &	\textbf{0.9027} &	\textbf{0.5751} &	\textbf{0.7216}	 & 	 \textbf{0.7777} & \textbf{0.2779}	& 	 \textbf{0.4457} \\
    \bottomrule
\end{tabular}}%

    \label{tab:ablation}
\end{table}

\begin{table}[!t]
\centering
    \caption{Ablation study with DeepFM as the base model.}
    \small
{\setlength{\tabcolsep}{0.3em}\begin{tabular}{l cccccc}
\toprule
 &\multicolumn{3}{c}{ Amazon } &\multicolumn{3}{c}{ Alipay } \\
\midrule
Model
   & AUC & NDCG & Recall & AUC & NDCG & Recall \\
    \midrule
	 DeepFM (Base) &	 0.8941& 0.5416 & 0.6877 & 0.8018 & 0.2501 & 0.4177 \\
	 \midrule
	 Ho Teachers &	 0.8550 &0.4371 & 0.5817&0.7797 &0.2669 &0.4265\\
		w.o. FDA &	 0.8860& 0.5528& 0.6953& 0.7811&0.2817 &0.4444 \\
	CausalD & \textbf{0.8948} & \textbf{0.5638}&\textbf{0.7016} & \textbf{0.8031}& \textbf{0.3053}& \textbf{0.4657} \\
    \bottomrule
\end{tabular}}%

    \label{tab:ablation_dfm}
\end{table}

%
%
%

\begin{table}[!t]
\centering
    \caption{Recommendation performance of CausalD as changing the number of heterogeneous teachers.}
   \small
{\setlength{\tabcolsep}{0.4em}\begin{tabular}{l cccccc}
\toprule
 &\multicolumn{3}{c}{ Amazon } &\multicolumn{3}{c}{ Alipay } \\
\midrule
\#Teachers
   & AUC & NDCG & Recall & AUC & NDCG & Recall \\
    \midrule
	 2 &	 0.8910 &	 0.5304 &	 0.6806 & 	 \textbf{0.7866} & 	 0.2682 & 	 0.4171 \\
	 4 &	 0.8869 &	 0.5429 &	 0.6883 & 	 0.7775 & 	 0.2812 & 	 0.4351 \\
	 8 &	\textbf{0.8965} &	 \textbf{0.5652} &	\textbf{0.7114} & 	 0.7834 & 	 \textbf{0.2902} & 	0.4498 \\
	 16 &	 0.8931 &	 0.5615 &	 0.7062 & 	 0.7723 & 	0.2832 & 	 \textbf{0.4514} \\
    \bottomrule
\end{tabular}}%

    \label{tab:teachernum}
\end{table}

\subsection{In-depth Model Analysis (RQ3)} \label{sec:indepthAnalysis}

\vpara{Ablation Studies.} We are interested in whether the critical components in CausalD all contribute to its effectiveness. In this regard, we evaluate whether 1) heterogeneous teachers outperform homogeneous teachers; and 2) whether the back-door adjustment loss and the front-door adjustment loss improve the performance. Specifically,
we construct \textbf{\texttt{w.o. FDA}} which denotes CausalD trained without front-door adjustment loss. By removing the back-door adjustment loss from \texttt{w.o. FDA}, we obtain the \textbf{\texttt{Base}} (DIN) model. Model \textbf{\texttt{Ho Teachers}} means replacing the heterogeneous teachers in CausalD with homogeneous teachers, which are pre-trained on randomly split data.

According to the results shown in Table \ref{tab:ablation}, we observe that removing any of these three components (heterogeneous teachers, BDA loss, FDA loss) will incur a performance degradation. Notably, removing FDA loss (\texttt{CausalD} $\rightarrow$ \texttt{w.o FDA}) leads to NDCG $-19.5\%$ relatively on the AliPay dataset. We attribute the result to the effective capability of front-door adjustment in reducing confounding effects. The performance drop caused by removing BDA loss (\texttt{w.o FDA} $\rightarrow$ \texttt{Base}) demonstrates that the back-door adjustment loss is helpful for causal intervention. Not surprisingly, CausalD trained with homogeneous teachers achieves inferior results, which show sampling heterogeneous mediator values given the input is essential for FDA. These results jointly demonstrate the rationality of our framework design. \zsy{We have mostly similar observations with DeeFM as the base model, according to the results listed in Table \ref{tab:ablation_dfm}.}

%

\vpara{Effect of the Teacher Number.} To investigate how the number of teachers affects the performance, we consider some variants of CausalD trained with $\{ 2,4,8,16 \}$ teachers, respectively. Table \ref{tab:teachernum} summarizes the results. By jointly analyzing Table \ref{tab:teachernum} and Table \ref{tab:comparison_DIN}, we have the following findings:
\begin{itemize}[leftmargin=*]
	\item Increasing the number of teachers mostly leads to performance gain. A larger number of teachers means that CausalD conducts more comprehensive sampling on the counterfactual space and thus estimates the causal effect in Equation (\ref{eq:original_fd}) more accurately.
	\item When further increasing the number of teachers from 8 to 16, we observe little performance gain in many cases. We attribute this phenomenon to that teachers themselves might become less effective with smaller training datasets. In larger dataset Alipay, CausalD achieves the best Recall.
	\item When varying the number of teachers, CausalD consistently outperforms the base model DIN. These results demonstrate that the front-door adjustment and the causal distillation can facilitate recommendation.
\end{itemize}


\vpara{Effect of the Loss Coefficients.} CausalD introduces two losses, \ie, the back-door adjustment loss, and the front-door adjustment loss. To investigate how the loss coefficients ($\lambda_{BDA}$ and $\lambda_{FDA}$ in Equation (\ref{eq:loss})) affect the performance, we vary $\lambda_{BDA}$ and $\lambda_{FDA}$ among $\{0.01, 0.1, 1, 10\}$, and test the performance. According to Figure \ref{fig:lambda}, we can find that 1) the recommendation performance is mostly insensitive to these hyper-parameters. Directly leaving them as $\lambda_{BDA}=\lambda_{FDA}=1$ will achieve substantial improvements over the base model; 2) increasing $\lambda_{FDA}$ consistently leads to performance gains, which again shows the merits of front-door adjustment for recommender systems; and 3) increasing $\lambda_{BDA}$ are mostly helpful while relatively large $\lambda_{BDA}$ might be inferior. This is reasonable since $\mathcal{L}_{BDA}$ is an auxiliary loss for $\mathcal{L}_{FDA}$. Larger $\mathcal{L}_{BDA}$ might incur less model optimization on $\mathcal{L}_{FDA}$.


\section{Conclusion and Future Work} \label{sec:conclusion}

%
%

In this work, we study how to alleviate the performance heterogeneity problem in recommender systems. Though the \textit{natural source} of this problem, \ie, imbalance of training data distribution over users, is inevitable, we find that model training enlarges the heterogeneity, \ie, a \textit{model source}. By analyzing the recommendation data generation process with the causal graph, we find that there are unobserved confounders that mainly mislead the estimation $P(Y \mid X)$ of tail users. Different from existing works that require unbiased data or assume unconfoundedness, we handle unobserved confounders using FDA. We propose the causal multi-teacher distillation framework (CausalD) as an efficient approximation of FDA by distilling causal effects estimated by heterogeneous teachers into a recommendation model. We conduct experiments on three public datasets, validating the effectiveness of CausalD on improving recommendation and alleviating performance heterogeneity.

To the best of our knowledge, this work takes the initiative to incorporate FDA into recommendation models. FDA has the enticing merit to deal with observed confounders that cause the spurious correlation of historical behaviors and next behaviors. Bias is ubiquitous and various in kind, and we believe this new paradigm is general for alleviating model training biases in a unified framework. There are many research directions remaining for exploration. Apart from performance heterogeneity, there are more issues caused by biased model training, such as filter bubble. We are interested in whether CausalD can alleviate these issues in the future. Another future direction relates to performing FDA on causally discovered causal graphs. Lastly, we plan to extend CausalD to state-of-the-art recommendation models that leverage various content features and side information.

\ifCLASSOPTIONcompsoc
  \section*{Acknowledgments}
\else
  \section*{Acknowledgment}
\fi

\begin{sloppypar}
This work was supported in part by National Natural Science Foundation of China (62006207, U20A20387, 62037001), Key R \& D Projects of the Ministry of Science and Technology (2020YFC0832500), Young Elite Scientists Sponsorship Program by CAST (2021QNRC001), Zhejiang Province Natural Science Foundation (LR19F020006, LQ21F020020), Project by Shanghai AI Laboratory (P22KS00111), Program of Zhejiang Province Science and Technology (2022C01044), the StarryNight Science Fund of Zhejiang University Shanghai Institute for Advanced Study (SN-ZJU-SIAS-0010).
\end{sloppypar}


\bibliographystyle{IEEEtran}
\bibliography{sections/9.citations}

\ifCLASSOPTIONcaptionsoff
  \newpage
\fi

\begin{IEEEbiography}[{\includegraphics[width=1in,height=1.25in,clip,keepaspectratio]{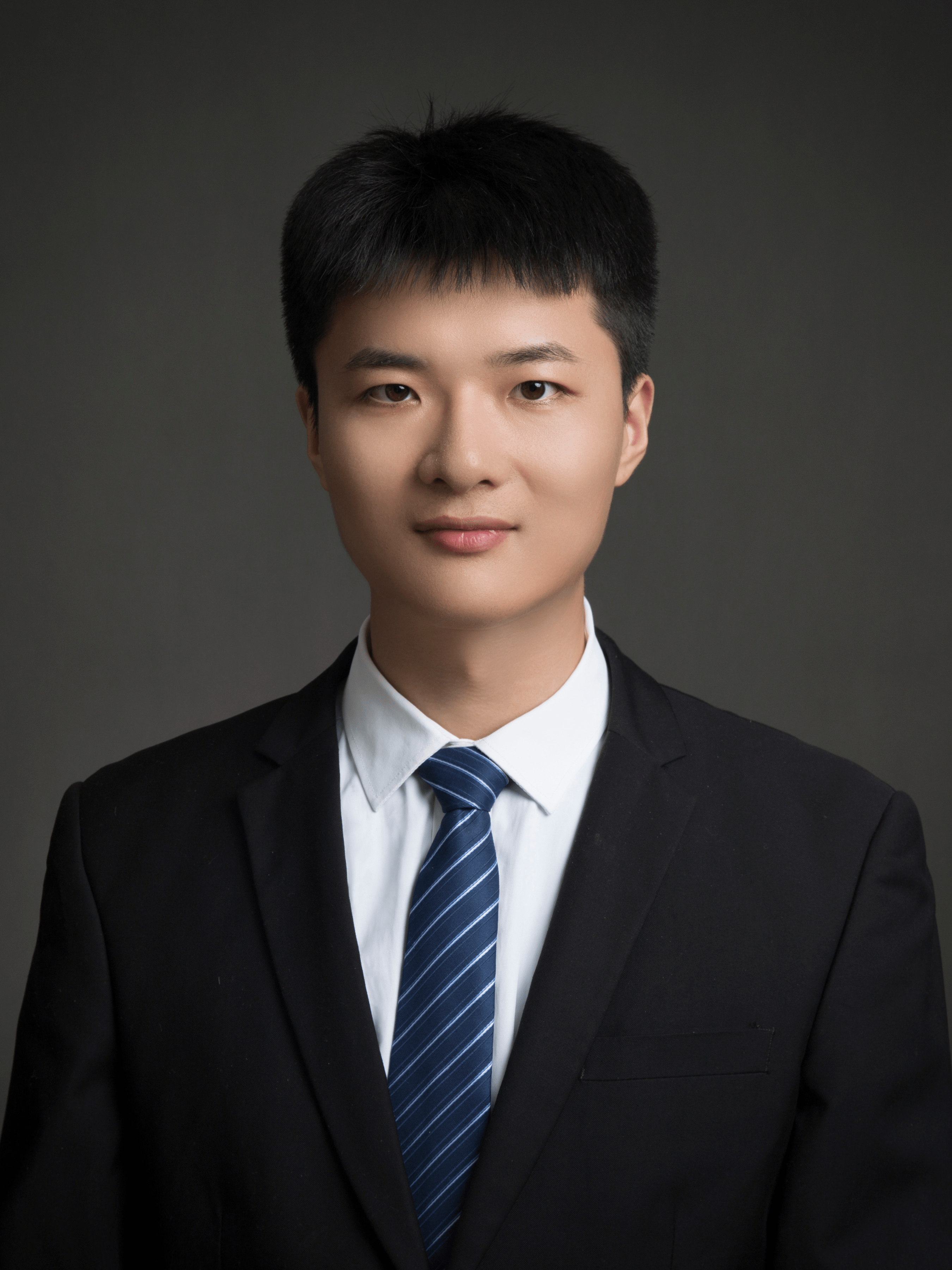}}]{Shengyu Zhang}
is currently a Ph.D. student at Zhejiang University, Hangzhou, China. He got the China Scholarship Council Fellowship in 2020 and visited National University of Singapore, Singapore, as a visiting research scholar from 2021 to 2022 (expected). His current research interests are in information retrieval, and causally reguarlized machine learning.  Until now, he has published more than 20 papers in major international journals and conferences such as SIGKDD, SIGIR, ACM MM, WWW, CVPR, AAAI, and TKDE, etc. He also serves as Reviewer of several high-level international journals such as TKDE, TCYB, TNNLS, etc.
\end{IEEEbiography}

\begin{IEEEbiography}[{\includegraphics[width=1in,height=1.25in,clip,keepaspectratio]{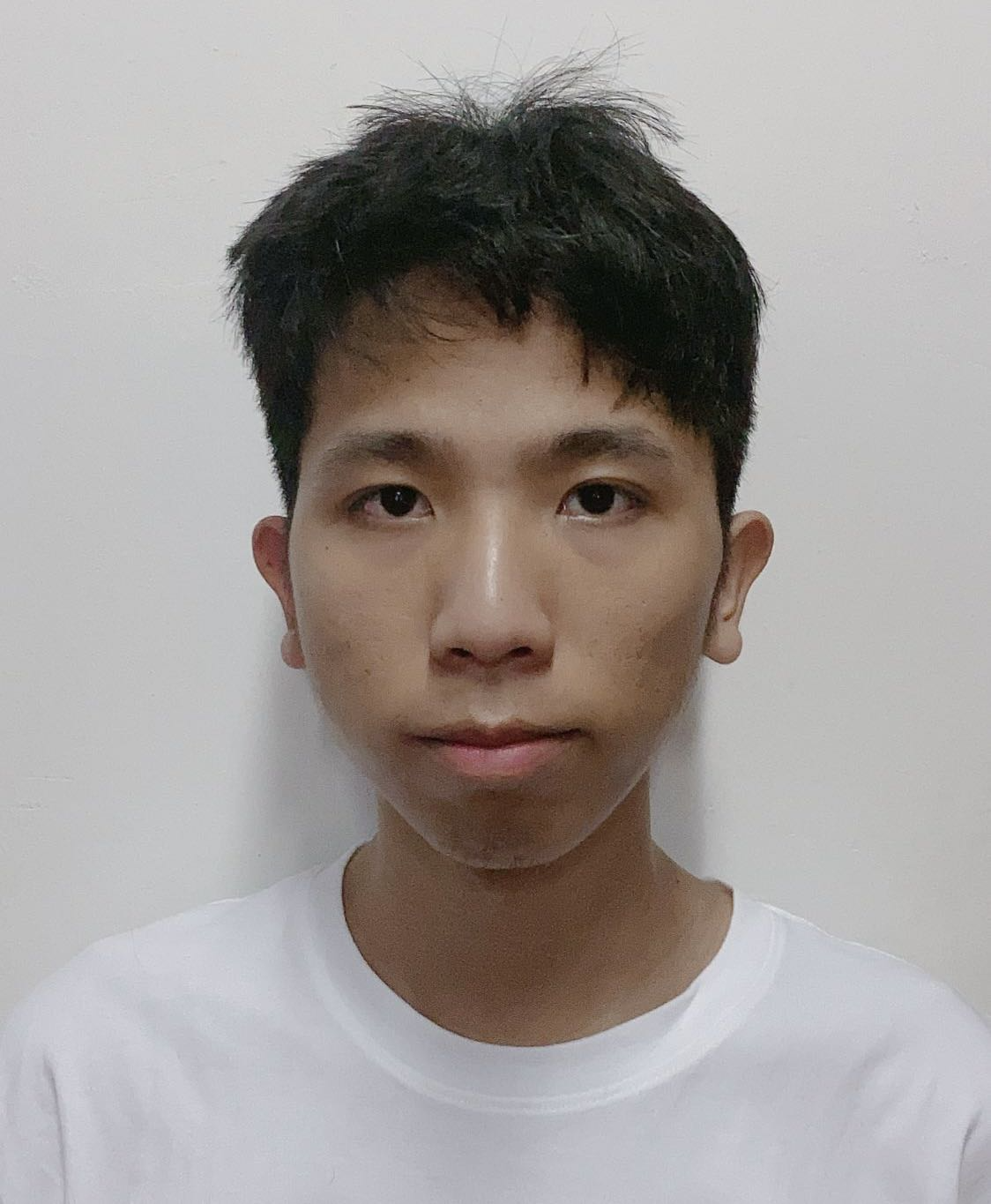}}]{Ziqi Jiang}
is currently an undergraduate student at the College of Computer Science, Zhejiang University, advised by Prof. Fei Wu.  His research interests lie at the intersection of multimedia, AI security, and causal inference, with the goal to build socially efficient and robust machine learning systems. He has published papers in ACM MM.
\end{IEEEbiography}

\begin{IEEEbiography}[{\includegraphics[width=1in,height=1.25in,clip,keepaspectratio]{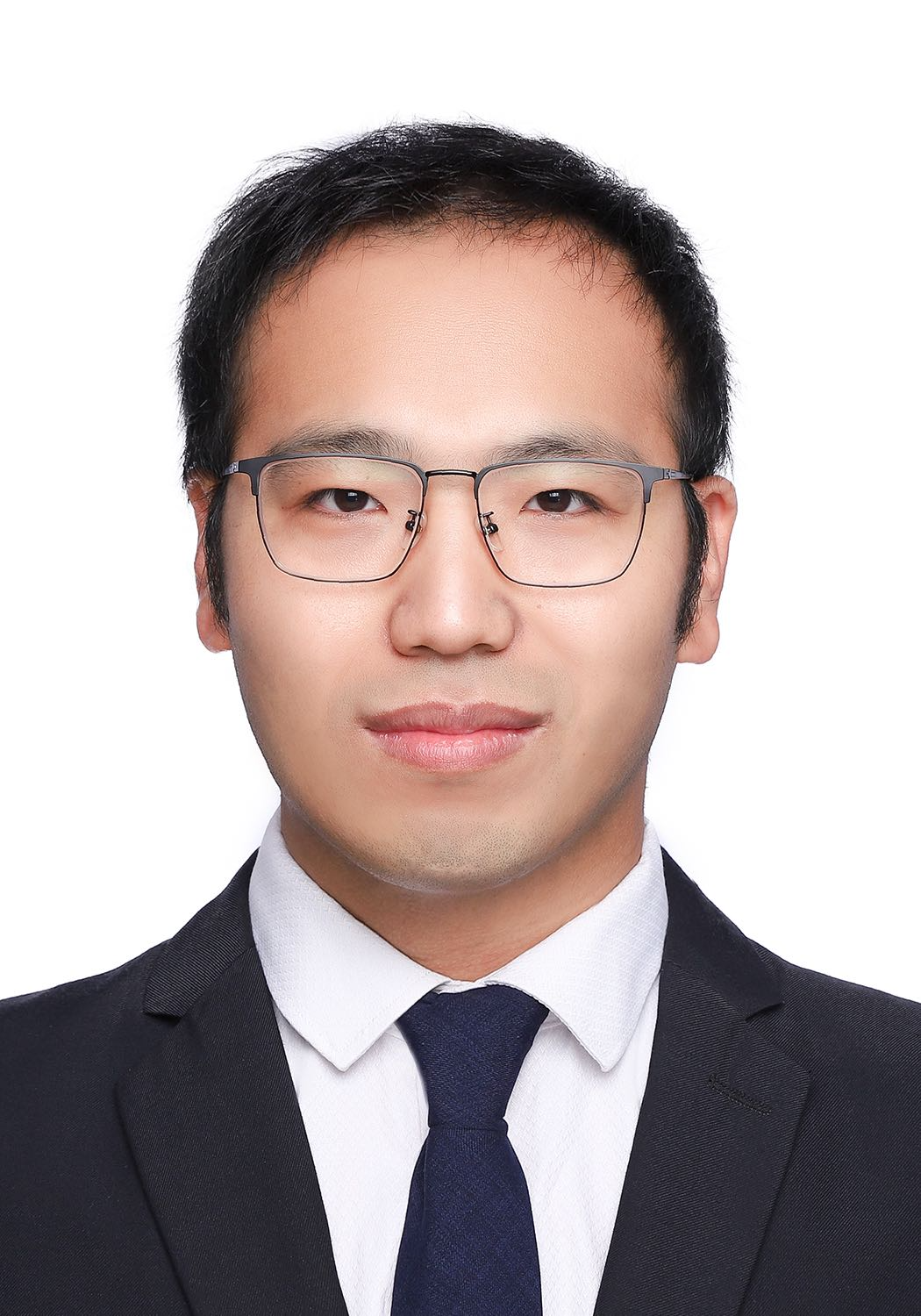}}]{Jiangchao Yao}
is an Assistant Professor at Shanghai Jiao Tong University. He got his dual Ph.D. degree under the supervision of Ya Zhang in Shanghai Jiao Tong University and Ivor W. Tsang in University of Technology Sydney in 2019. His research interests lie on Trustworthty Machine Learning and Reasoning, making machine learning robust, interpretable and scalable. He has published over 20 papers in major international journals and conferences such as ICML, ICLR, NeurIPS, KDD, TPAMI, TIP, TKDE, etc..
\end{IEEEbiography}

\begin{IEEEbiography}[{\includegraphics[width=1in,height=1.25in,clip,keepaspectratio]{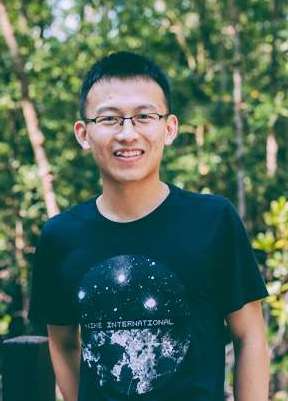}}]{Fuli Feng}
is a professor at the University of Science and Technology of China (USTC). He received Ph.D. in Computer Science from NUS in 2019. His research interests include information retrieval, data mining, and multi-media processing. He has over 30 publications appeared in several top conferences such as SIGIR, WWW, and MM, and journals including TKDE and TOIS. His work on Bayesian Personalized Ranking has received the Best Poster Award of WWW 2018. Moreover, he has been served as the PC member for several top conferences including SIGIR, WWW, WSDM, NeurIPS, AAAI, ACL, MM, and invited reviewer for prestigious journals such as TOIS, TKDE, TNNLS, TPAMI, and TMM.
\end{IEEEbiography}

\begin{IEEEbiography}[{\includegraphics[width=1in,height=1.25in,clip,keepaspectratio]{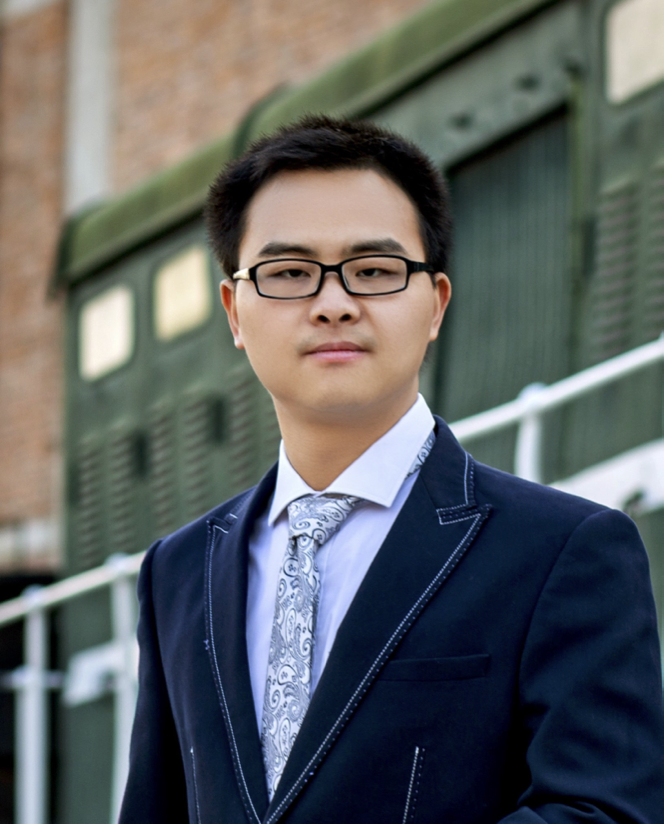}}]{Kun Kuang}
received his Ph.D. degree from Tsinghua University in 2019. He is now an Associate Professor in the College of Computer Science and Technology, Zhejiang University. He was a visiting scholar with Prof. Susan Athey's Group at Stanford University. His main research interests include Causal Inference, Artificial Intelligence, and Causally Regularized Machine Learning. He has published over 40 papers in major international journals and conferences, including SIGKDD, ICML, ACM MM, AAAI, IJCAI, TKDE, TKDD, Engineering, and ICDM, etc.
\end{IEEEbiography}

\begin{IEEEbiography}[{\includegraphics[width=1in,height=1.25in,clip,keepaspectratio]{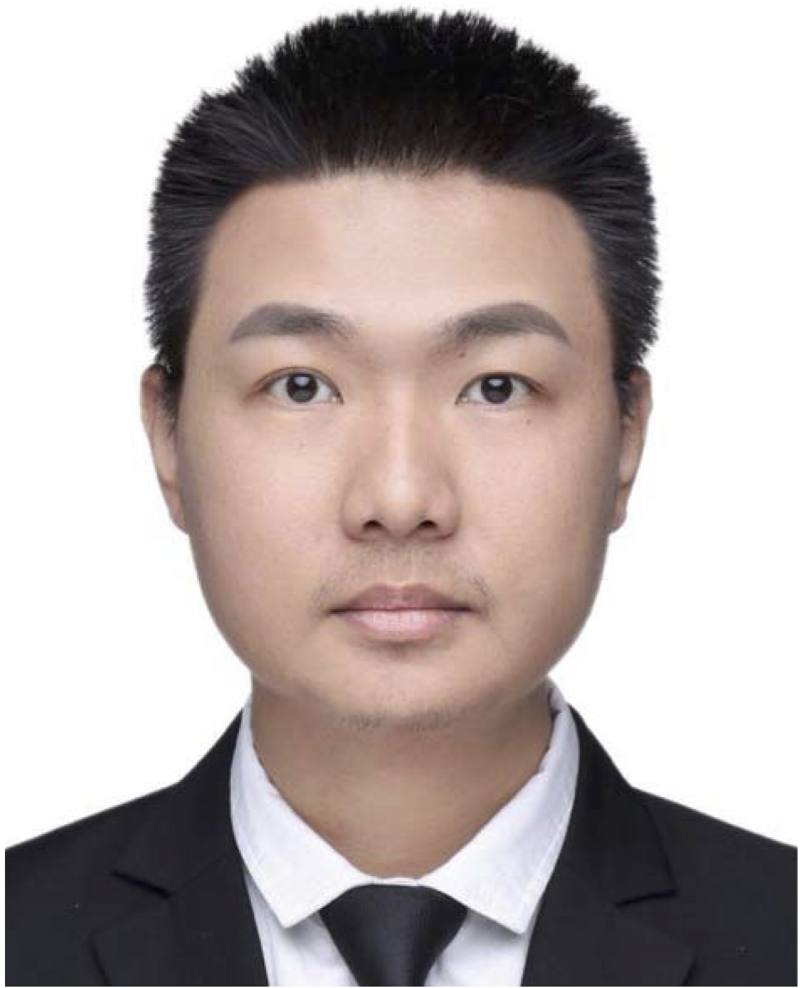}}]{Zhou Zhao}
(Member, IEEE) received the B.S.
and Ph.D. degrees in computer science from The
Hong Kong University of Science and Technology,
in 2010 and 2015, respectively. He is currently an
Associate Professor with the College of Computer
Science, Zhejiang University. His research interests
include machine learning and data mining.
\end{IEEEbiography}

\begin{IEEEbiography}[{\includegraphics[width=1in,height=1.25in,clip,keepaspectratio]{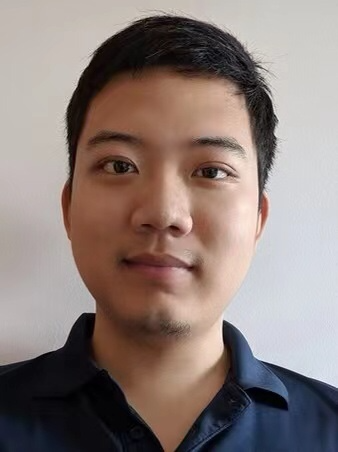}}]{Shuo Li} is an undergraduate at School of Computing, National University of Singapore (NUS), and works as an research intern at Lazada data science team. His research interests include multi-modal learning and natural language processing.
\end{IEEEbiography}

\begin{IEEEbiography}[{\includegraphics[width=1in,height=1in,clip,keepaspectratio]{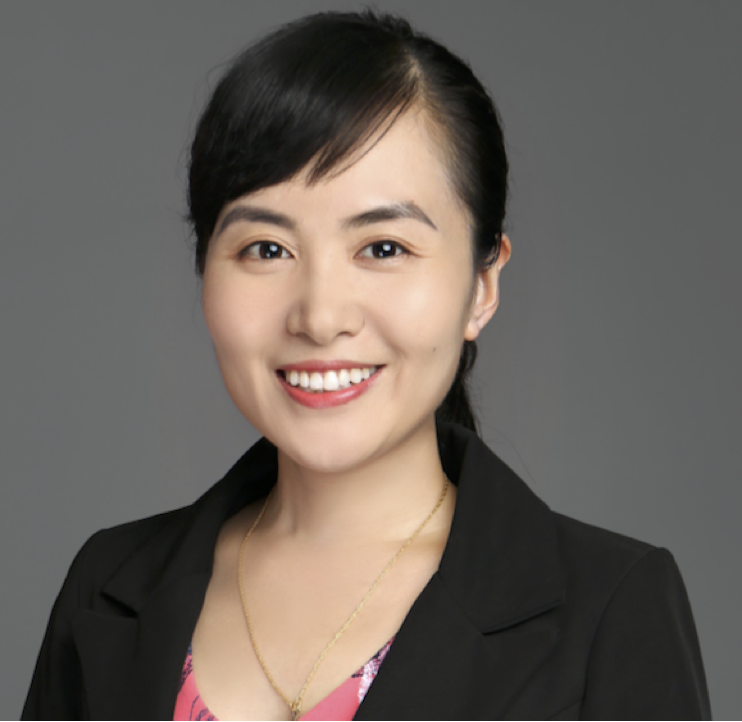}}]{Hongxia Yang} received the PhD degree in statistics from Duke University, in 2010. She is working
as a senior staff data scientist and director with
the Alibaba Group. She has published more than
60 papers and held nine filed/to be filed US patents and is serving as the associate editor of
Applied Stochastic Models in Business and Industry.
\end{IEEEbiography}

\begin{IEEEbiography}[{\includegraphics[width=1in,height=1.25in,clip,keepaspectratio]{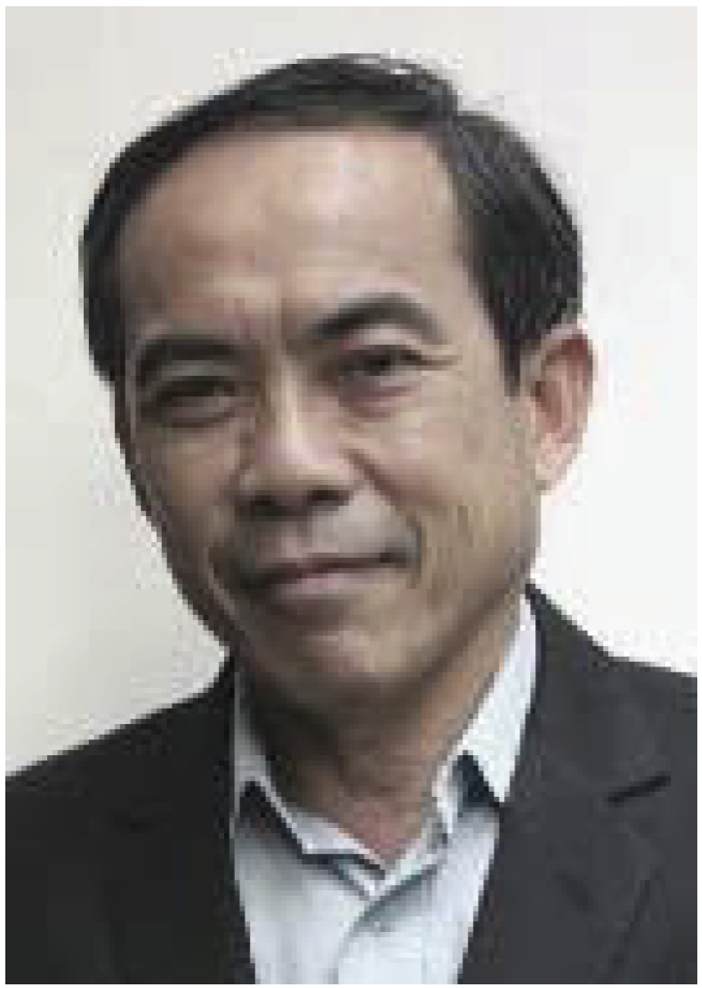}}]{Tat-Seng Chua} is the KITHCT Chair Professor
at the School of Computing, National University
of Singapore. He was the Acting and Founding
Dean of the School during 1998-2000. Dr
Chua‘s main research interest is in multimedia
information retrieval and social media analytics.
In particular, his research focuses on the
extraction, retrieval and question-answering
(QA) of text and rich media arising from the
Web and multiple social networks. He is the
co-Director of NExT, a joint Center between
NUS and Tsinghua University to develop technologies for live social
media search. Dr Chua is the 2015 winner of the prestigious ACM
SIGMM award for Outstanding Technical Contributions to Multimedia
Computing, Communications and Applications. He is the Chair of
steering committee of ACM International Conference on Multimedia
Retrieval (ICMR) and Multimedia Modeling (MMM) conference series. Dr
Chua is also the General Co-Chair of ACM Multimedia 2005, ACM CIVR
(now ACM ICMR) 2005, ACM SIGIR 2008, and ACM Web Science 2015.
He serves in the editorial boards of four international journals. Dr. Chua
is the co-Founder of two technology startup companies in Singapore.
He holds a PhD from the University of Leeds, UK.
\end{IEEEbiography}

\begin{IEEEbiography}[{\includegraphics[width=1in,height=1.25in,clip,keepaspectratio]{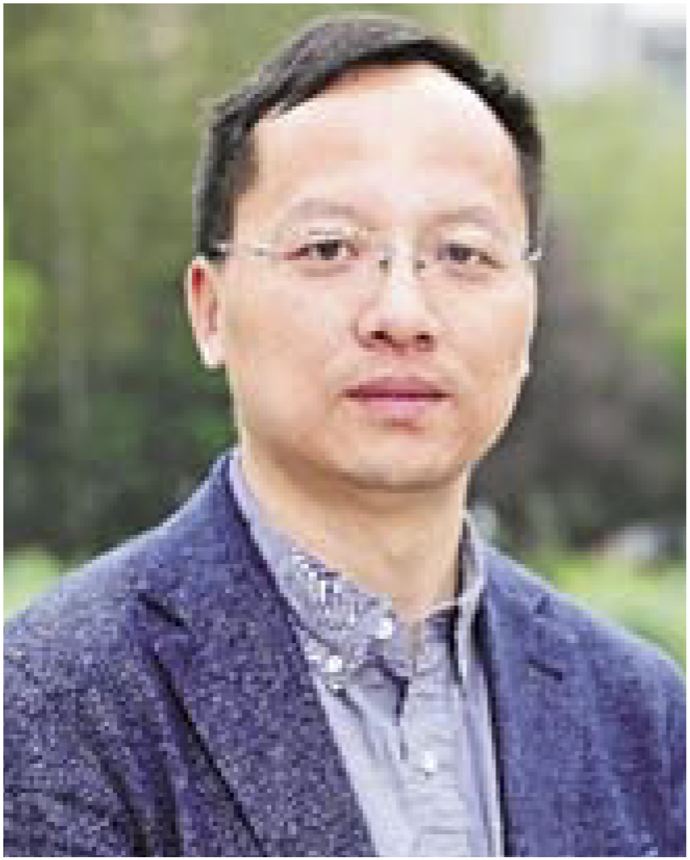}}]{Fei Wu}
(Senior Member, IEEE) received the Ph.D. degree from Zhejiang
University, Hangzhou, China. He was a Visiting
Scholar with the Prof. B. Yu’s Group, University of
California at Berkeley, Berkeley, from 2009 to 2010.
He is currently a Full Professor with the College of
Computer Science and Technology, Zhejiang University. His current research interests include multimedia retrieval, sparse representation, and machine
learning.
\end{IEEEbiography}

\end{document}